\documentstyle[preprint,aps,epsf,cite,tighten]{revtex}

\def\beq{\begin{equation}}   
\def\eeq{\end{equation}}

\newcommand{\nc}{\newcommand}
\nc{\grad}{\nabla}
\nc{\tr}{\mathop{\rm tr}}
\nc{\half}{{1\over 2}}
\nc{\third}{{1\over 3}}
\nc{\be}{\begin{equation}}
\nc{\ee}{\end{equation}}
\nc{\bea}{\begin{eqnarray}}
\nc{\eea}{\end{eqnarray}}

\def\Tr{{\rm Tr}}
\nc{\dint}[2]{\int\limits_{#1}^{#2}}
\nc{\D}{\displaystyle}
\nc{\PDT}[1]{\frac{\partial #1}{\partial t}}
\nc{\tw}{\tilde{w}}
\nc{\tg}{\tilde{g}}
\nc{\newcaption}[1]{\centerline{\parbox{5.6in}{\caption{\footnotesize{#1
}}}}}
\def\href#1#2{#2}

\nc{\al}{\alpha}
\nc{\ga}{\gamma}
\nc{\de}{\delta}
\nc{\ep}{\epsilon}
\nc{\ze}{\zeta}
\nc{\et}{\eta}
\renewcommand{\th}{\theta}
\nc{\Th}{\Theta}
\nc{\ka}{\kappa}
\nc{\la}{\lambda}
\nc{\rh}{\rho}
\nc{\si}{\sigma}
\nc{\ta}{\tau}
\nc{\up}{\upsilon}
\nc{\ph}{\phi}
\nc{\ch}{\chi}
\nc{\ps}{\psi}
\nc{\om}{\omega}
\nc{\Ga}{\Gamma}
\nc{\De}{\Delta}
\nc{\La}{\Lambda}
\nc{\Si}{\Sigma}
\nc{\Up}{\Upsilon}
\nc{\Ph}{\Phi}
\nc{\Ps}{\Psi}
\nc{\Om}{\Omega}
\nc{\ptl}{\partial}
\nc{\del}{\nabla}
\nc{\ov}{\overline}
\nc{\gsl}{\!\not}
\nc{\bi}[1]{\bibitem{#1}}
\nc{\fr}[2]{\frac{#1}{#2}}
\nc{\gm}{\mbox{$\gamma_{\mu}$}}
\nc{\gn}{\mbox{$\gamma_{\nu}$}}
\nc{\Le}{\mbox{$\fr{1+\gamma_5}{2}$}}
\nc{\Ri}{\mbox{$\fr{1-\gamma_5}{2}$}}
\nc{\GD}{\mbox{$\tilde{G}$}}
\nc{\gf}{\mbox{$\gamma_{5}$}}
\nc{\Ima}{\mbox{Im}}
\nc{\Rea}{\mbox{Re}}

\nc{\av}{\langle \ph\rangle}
\nc{\ntwo}{${\cal N}\!\!=\!2\;$}
\nc{\none}{${\cal N}\!\!=\!1\;$}
\nc{\nfour}{${\cal N}\!\!=\!4\;$}

\newcommand{\bm}[1]{{\mbox{\boldmath $#1$}}}

\def \bi{\bibitem}
\nc{\rf}[1]{(\ref{#1})}
\def \del{\partial}

\begin{document}
\draft
\preprint{
\vbox{\hbox{TPI-MINN-00/02}
      \hbox{UMN-TH-1835/00}
      \hbox{hep-th/0006028}}}
\title{
Marginal Stability
and the Metamorphosis of  BPS States
}
\author{
Adam Ritz$^{\,a}$, Mikhail Shifman$^{\,b}$,
  Arkady Vainshtein$^{\,b}$, and  Mikhail Voloshin$^{\,b,c}$
}
\address{
$^a$Department of Applied Mathematics and Theoretical
Physics,
  Centre for Mathematical Sciences, University of Cambridge,
Wilberforce Rd., Cambridge CB3 0WA, UK\\
$^b$Theoretical Physics Institute, University of Minnesota,
116 Church St SE,\\ Minneapolis, MN 55455
\\
$^c$Institute of Experimental and Theoretical Physics,
B. Cheremushkinskaya 25,\\ Moscow
117259, Russia
}
\maketitle
\thispagestyle{empty}
\setcounter{page}{0}
\begin{abstract}
We discuss the restructuring of the BPS 
spectrum which occurs on certain submanifolds of the moduli/parameter
space -- the curves of the marginal stability (CMS) --
using quasiclassical methods.  We argue that in general
a `composite' BPS soliton swells in coordinate space as one approaches
the CMS and that, as a bound state of two `primary' solitons,
its dynamics in this region is determined by
non-relativistic supersymmetric quantum mechanics.
Near the CMS the bound state has a wave function which is 
highly  spread out. Precisely on the CMS the
bound state level reaches the continuum, the composite state
delocalizes in coordinate space, and restructuring of the
spectrum can occur. We present a detailed analysis of this behavior
in a two-dimensional ${\cal N}\!\!=\!2$ Wess-Zumino model with two
chiral fields, and then discuss how it arises  in the context of
`composite' dyons near weak coupling CMS curves in \ntwo
supersymmetric gauge theories.
We also consider cases where some states become massless
 on the CMS.
\end{abstract}

\newpage

\tableofcontents

\newpage
\section{Introduction}
\label{sec:intro}

Centrally extended supersymmetry algebras admit a special class of
massive representations which preserve some fraction of the
supersymmetry of the vacuum, and consequently form multiplets which
are smaller than a generic massive representation. The states lying
in these shortened (or BPS, after Bogomol'nyi-Prasad-Sommerfield)
multiplets are extremely useful probes of the theory because 
on one hand their spectrum is determined almost entirely by
kinematical constraints (i.e. by the central charges), while on the
other the multiplet structure ensures their generic stability. More
precisely, the fact that BPS states lie in short multiplets implies
that they must remain BPS, unless a degeneracy of several BPS 
multiplets is achieved which can then combine to
form a generic massive multiplet. In the absence of such an exotic
scenario, the dynamics of the BPS sector forms a closed subsystem.

The stability of BPS (particle) states follows from the fact that
their masses are determined by the superalgebra
to be the expectation values of the central charge(s) ${\cal Z}_i$. 
Since the central charges ${\cal Z}_i$ are additive, this implies 
via the triangle inequality that a BPS state whose mass is
\be
 M=\Big|\Big\langle\sum_i {\cal Z}_i\Big\rangle\Big|
\ee
is stable with respect
to decay into BPS `constituents' with 
masses $M_i\!=\!| \langle {\cal Z}_i\rangle|$,
\be
 M \leq \sum_i M_i. \label{ineq}
\ee
Even at points where this equality is saturated there is no phase
space for a physical decay. Thus one concludes that BPS particles 
are indeed stable.

However, restructuring of the spectrum is nonetheless possible because
of the existence of special submanifolds of the moduli/parameter space
where the inequality (\ref{ineq}) is saturated.  Specifically, this
allows for discontinuities of the spectrum
with respect to changes in these moduli. Such changes are `unphysical'
in the sense that one shifts between different superselection sectors.
Nonetheless, one is often interested in considering such an evolution,
as it may correspond to the extrapolation from a weakly coupled to a
strongly coupled regime. In this case, the stability of BPS states
can often be used to infer information about the strongly coupled region.
The caveat of course is that one should not cross a submanifold where
the bound (\ref{ineq}) is saturated, and where restructuring of the spectrum 
may occur and BPS states may for example disappear. Such 
submanifolds are consequently known as 
curves of marginal stability (CMS), although their actual co-dimension in the
moduli/parameter space will vary.

Marginal stability curves, and the corresponding discontinuities
of the BPS spectrum, are quite ubiquitous in theories
with centrally extended supersymmetry algebras. Examples include:
the existence of a CMS for the BPS
soliton spectrum in general classes of two dimensional 
models discussed by Cecotti and Vafa \cite{Vafa}; and the CMS
for the BPS particle spectrum in \ntwo supersymmetric gauge
theories \cite{sw,n2cms,fb1,bf2}. In the latter case an explicit 
demonstration of the discontinuity of the spectrum across these
curves in the vacuum moduli space was provided in generic SU(2) theories by
Bilal and Ferrari \cite{fb1,bf2}. The realization of these dyonic
states in terms of Type IIB string junctions has also
led to the appearance of marginal stability conditions in this 
context \cite{bergman,Wsj}. Furthermore, a discontinuity in the BPS spectrum
of wall solutions in a Wess-Zumino model with the 
Taylor-Veneziano-Yankielowicz superpotential \cite{tvy}, which
leads to a glued potential, was also observed recently by Smilga and
Veselov \cite{SmVe,Sm}. The discontinuity arises in this case as a
function of the mass parameter -- a feature also observed
in some of the models to be considered in this paper. 
Finally, we also mention that
marginal stability conditions have more recently been studied
in the context of string compactification on manifolds with nontrivial
cycles \cite{cyau}.

\begin{figure}[ht]
\epsfxsize=5cm
 \centerline{%
   \epsfbox{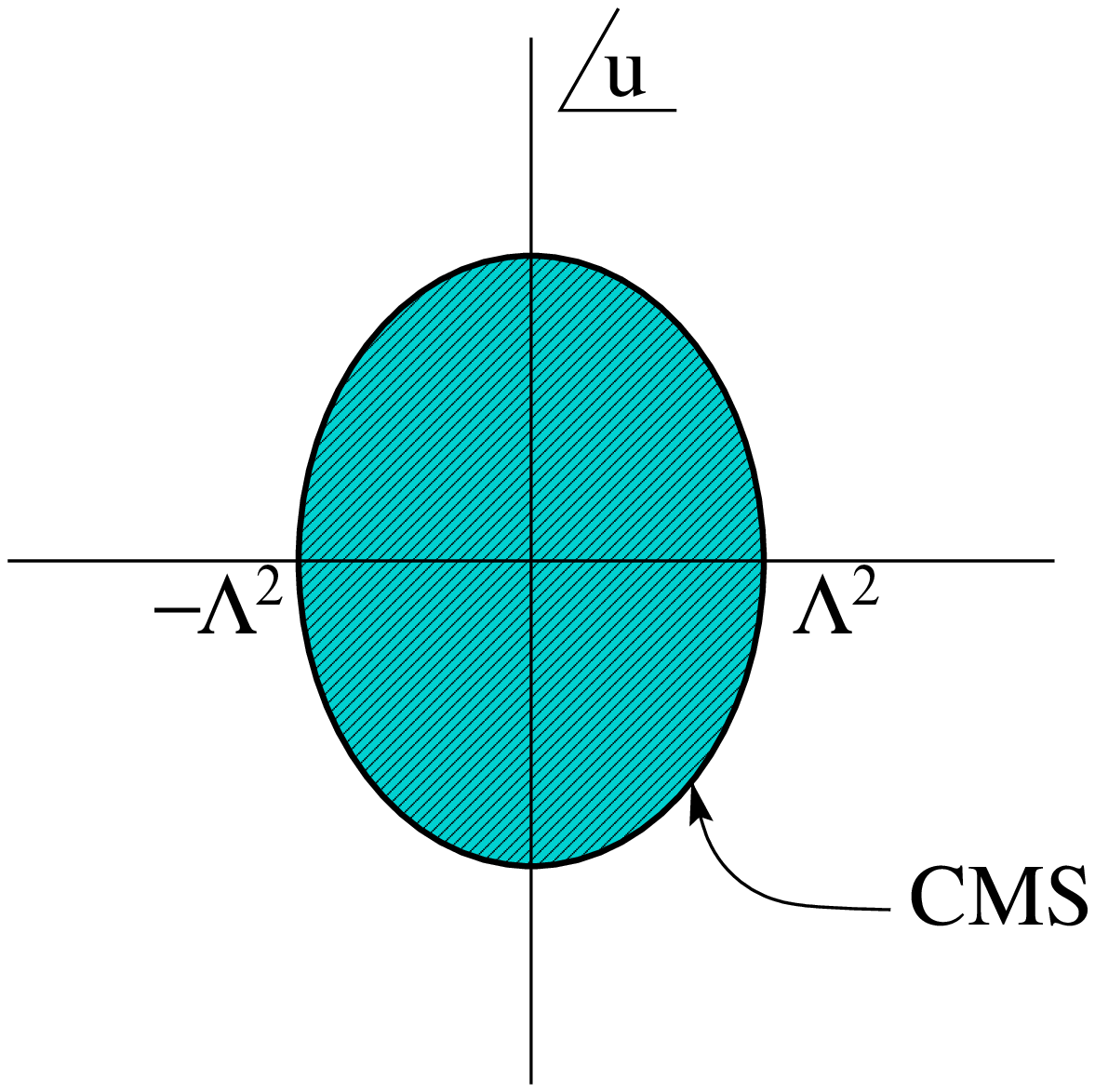}%
         }
 \newcaption{A schematic representation of the moduli space for \ntwo
SYM with gauge group SU(2) in terms of the vev
$u=\langle{\rm tr}\,\ph^2\rangle$ of the adjoint scalar
$\ph$. The $W$ bosons only exist outside the shaded region, which consequently
determines their stability domain.}
\label{fig:Wb}
\end{figure}

In order to illustrate the discussion with a particular example, 
we recall that the notion of marginal stability arises, in particular, in the
Seiberg-Witten solution \cite{sw} of ${\cal N}\!\!=\!2$ supersymmetric
(SUSY) gauge theories (see e.g. \cite{fb1}). In the simplest 
example of pure super Yang-Mills with gauge group SU(2), there 
is a one dimensional elliptical curve of marginal stability in the 
moduli space (see Fig.~\ref{fig:Wb}).
\firstfigfalse
On crossing this curve by varying the moduli a
restructuring of the spectrum of BPS states takes place. For
instance, the electrically charged vector bosons $W^\pm$ only 
exist outside the CMS, and disappear from the spectrum in the interior
region. To make these notions a little more general, we can 
define a ``stability domain'' as a submanifold of the moduli space 
in which a particular BPS state exists. This domain will always 
be bounded by a CMS.
In this example, the $W$ boson has a stability domain in the
exterior region illustrated in Fig.~\ref{fig:Wb}. On crossing the CMS
from the stability domain, it is usually stated that the $W$ bosons 
``decay'' into a two particle state consisting of a monopole and 
a dyon with unit electric charge. This interpretation 
is a little awkward because for a particle to properly decay it must exist 
in the spectrum, at least as a quasi-stationary state, and this is not 
true after crossing the CMS. The question then arises as to 
exactly what happens on the CMS resulting in
the apparent discontinuity of the BPS spectrum.

In this paper we will suggest a physical interpretation of this
phenomenon, which we summarize below. For this purpose, its  
convenient to continue with the $W$ boson example to make the 
ideas more concrete. However, one should bear in mind that 
this system is not directly accessible to the
semi-classical techniques that are used in this paper, because the CMS curve 
lies at strong coupling. Nonetheless, we will argue that there are
several constraints ensuring, at least qualitatively, the generality of this
behavior. 

Specifically, the emerging picture is that when the 
moduli approach the CMS, the $W^\pm$ states swell in coordinate space
as they become more weakly bound. Near the CMS, but
still within the stability domain, one can interpret the $W^\pm$
as a composite particle built from two primary constituents
(a monopole and a dyon of electric charge one), whose
interaction can be described by a nonrelativistic (super)potential
depending on the relative separation,
within the framework of  supersymmetric  quantum mechanics
(SQM) \cite{EWi}. As one approaches the CMS, the  separation
of the two primary constituents diverges, while the bound state level 
reaches the continuum (i.e. the binding energy vanishes). Further
motion after crossing the CMS leads to an SQM superpotential which
fails to exhibit a supersymmetric ground state separated by a gap 
from the continuum. Consequently, the ground state in the sector 
with unit electric charge is no longer the one-particle $W$ boson
state but rather a set of non-BPS two-particle  states 
forming a `long' multiplet.

We will argue that this picture is the general situation
for CMS curves associated with BPS particle states.
Namely, whenever a discontinuity occurs in the BPS spectrum
at a point in the parameter space, then certain
BPS states delocalize in coordinate space. Indeed,
the phenomenon of marginal stability of BPS states involves,
by definition, the alignment of central charges of primary states
in such a way that the binding energy vanishes. In this context it is
quite natural that crossing the CMS involves infrared effects, and the
`size' of the marginally stable state should diverge as the CMS is
approached. 
However, within this general picture of {\it delocalization}
one can identify several different mechanisms underlying this
behavior. 

The features are somewhat dimension dependent, so its
convenient to focus first on 1+1D which will be our primary concern
in this paper. We will then remark on certain aspects which
distinguish the behavior in 3+1D in particular.
Moreover, for solitons in 1+1D its convenient to distinguish two 
delocalization mechanisms.

\bigskip
(1) The first is when there are no massless fields
relevant to the problem, and consequently one can describe the interactions 
of the primary constituents using non-relativistic collective 
coordinate dynamics with linearly realized supersymmetry and short
range potentials. For a large
class of systems (including the ones to be considered here), its
possible to limit attention to just one collective coordinate -- the
relative separation $r$ of the primary solitons. 
We then observe two characteristic dynamical scenarios for the 
near CMS dynamics:
\begin{itemize}
\item In the first case, the short range potential is of deuteron-type
which remains finite on the CMS but possesses a single bound state,
whose wavefunction spreads out as the CMS is approached, while
the level reaches the continuum at this point. On crossing the CMS,
there is no longer a supersymmetric ground state reflecting the fact
that the BPS state has disappeared from the spectrum to be replaced as 
a ground state
by a non-BPS two-particle state with the same quantum numbers.
\item In the second scenario, the relative separation becomes an exact
modulus on the CMS, and the potential therefore vanishes at this
point. In this case, the composite state still delocalizes as the CMS
is approached as the wavefunction becomes highly spread out. The state
is however highly quantum mechanical and has no classical analogue. In
this case, we also observe that the potential may support (in general
different) composite states on each side of the CMS.
\end{itemize}
We will study a two-field model exhibiting both these dynamical
regimes in subsequent sections.

\bigskip
(2) The second delocalization mechanism arises when there are massless
fields involved, these being either the primary states themselves, or the
fields via which their interactions are mediated. 
\begin{itemize}
\item In situations where massive primary states interact via massless
exchange, we shall argue in Sec.~2 that attractive 
Coulomb-like interactions between the primary constituents must 
vanish on the CMS as a consequence of the structure of the BPS mass spectrum. 
\item The second mechanism arises when one or more of the
primary states is massless. This scenario may be taken as a special
case of (1) in that massless points arise generically as co-dimension
one submanifolds on the CMS curve. In this case it is 
not possible to reduce the effective dynamics to non-relativistic
quantum mechanics, and one must consider the effective theory
of the massless state. We note that more exotic examples of this 
behavior may arise (in higher dimensions)
at Argyres-Douglas points \cite{ad} in \ntwo SYM, or more
generally at second order critical points in supersymmetric theories.
\end{itemize}

Although we have framed this discussion in the context of 1+1D field
theories many of the features apply also in higher dimensions.
In particular, restructuring of the BPS spectrum via delocalization 
is apparently a generic phenomenon. However, an important 
distinction between 1+1D and, say, 3+1D is that in 1+1D an arbitrarily
small attraction is sufficient to form a bound state while 
in 3+1D this is not the case. For this reason long range forces 
play a special role in 3+1D, where Coulomb-like attractive potentials 
are required to form bound states at an
arbitrarily small effective coupling. We will discuss this
scenario in the form of composite dyons in \ntwo SYM. 
The general arguments outlined above will
be presented in Sec.~\ref{sec:two}, while particular examples of different
scenarios will be discussed in subsequent sections.

In this paper we focus first on class (1) and present an exhaustive
study of a particular two dimensional Wess Zumino model \cite{WZ} with 
\ntwo supersymmetry of the type considered previously
\cite{Shifman:1998wg} 
in a related context. 
This is a simple model which exhibits composite solitons (kinks) 
and a corresponding CMS curve accessible to
quasiclassical techniques. Thus it serves as an ideal arena to study
in detail the effective SQM which determines the presence or otherwise
of the composite soliton. The model involves two weakly interacting
chiral fields. In the decoupling limit there are `primary' BPS kinks for each
field, which when quantized lead to short BPS multiplets containing
one bosonic and one fermionic state (plus antiparticles). There
are also `composite' solitons which are combinations of the primary
configurations.

Switching on an interaction between the fields we see that the 
primary BPS solitons exist throughout the parameter space, while 
the composite solitons exhibit a finite stability domain bounded by
the CMS. We analyze in detail the effective SQM which exhibits the
composite soliton as a supersymmetric bound state, and verify  
the behavior described earlier with regard to the approach to
the CMS. Its worth noting that in this model the structure of the
stability domain is quite complex. In particular, there are different
dynamical regimes depending on which part of the CMS curve is crossed.
In most cases, a composite state only exists on one side of the CMS.
However, there are regions in the parameter space
where stability domains for particular
composite states meet on a CMS, and consequently (in general different)
composite states can exist on either side. In this latter regime we
observe that the relative separation of the primary states becomes
a modulus on the CMS, as the potential vanishes. 
Therefore this model exhibits both scenarios outlined in (1) above. 

The advantage of dealing with the Wess-Zumino model is that all the 
features of the non-relativistic quantum dynamics can be calculated 
analytically. In the vicinity of the CMS we obtain the explicit form for the 
(super)potential describing the interaction of the primary solitons
and are able to track 
the form of the bound state wavefunction right onto the CMS. 
Moreover, certain qualitative aspects
are apparently rather model independent due to the constraints imposed
by the BPS spectrum. To investigate the situation in 3+1D
we consider explicitly \ntwo SYM with gauge group SU(3) which 
contains a spectrum of primary and composite monopole (dyon) solutions.  
The two `primary' monopole solutions are embedded along each 
of the simple roots of the algebra. An embedding along the 
additional positive root leads to a
`composite' dyon which becomes marginally stable on a CMS which is
present in the semiclassical region. The major difference 
between the monopole case
in four dimensions in comparison to the two-field model in two
dimensions is the presence of massless exchanges resulting in a long
range Coulomb-like interaction, which can lead to bound states as noted above.
Recently there has been considerable interest in this
system \cite{ly,bhlms,tong,blly,bly,bl,n2quant}, in part because
the composite dyon configuration is an example of a
1/4-BPS state in ${\cal N}\!\!=\!4$ SYM. This work, which has
centered on the moduli space formulation of the low energy dynamics, 
has resulted in the detailed form of the long range interaction. 
We observe that, in accord
with the general expectations of Sec.~\ref{sec:two}, the attractive
component of the long range force (the term $\propto 1/r$ in the
effective  potential) vanishes on the CMS, while a repulsive component
($\propto 1/r^2$) remains. There is no attraction on the other side of
the CMS, the term $\propto 1/r$ change its sign. Thus, a BPS bound
state which exists in the stability domain on one side of the CMS becomes
more and more delocalized when approaching the CMS, and ceases to exist
on the other side. Accounting for the fact that long range forces are now
crucial, we observe that the qualitative picture is nonetheless quite
similar to the two-field Wess-Zumino model, in that the composite
state delocalizes on approach to the CMS.

The layout of the paper is as follows: 
In Sec.~\ref{sec:two} we present some general arguments 
constraining the dynamics of primary solitons near a CMS. 
Using these results we discuss, in a simplified
setting, the underlying mechanism involved in restructuring the spectrum,
introducing the necessary notation and definitions in passing. 
In this section, we also consider the embedding of the effective SQM
superalgebra within the superalgebra of the field theory. As a
specific example, we consider the realization of
the \ntwo superalgebra with central charges in two dimensions 
in the two soliton sector. In this regard its worth remarking that
this embedding shows explicitly how the presence of
field theoretic central charges is crucial in allowing a linear
realization of supersymmetry in the effective non-relativistic
dynamics.

Sec.~\ref{sec:three} presents a detailed analysis of the
(quasiclassical) solitons in the two-field Wess Zumino model with regard to
their BPS properties. We calculate the form of the CMS, and prove 
that outside the stability domain the BPS solution corresponding to 
the composite state ceases to exist. In
Sec.~\ref{sec:four} we derive and discuss the SQM which describes 
the interaction
between the primary solitons in the vicinity of CMS and determines whether or
not a supersymmetric bound state exists. We obtain analytic solutions
for the superpotential and the bound state wavefunction.

In Sec.~\ref{sec:five} we consider the more complex situation of monopoles and
dyons in \mbox{\ntwo} SYM with gauge group SU(3), and
review the form of the long range potential near the 
CMS \cite{ly,bhlms,tong,blly,bly,bl,n2quant}.
The attractive component vanishes on the CMS, in agreement with the 
general arguments of Sec.~\ref{sec:two}, while a repulsive component
remains leading to delocalization on the CMS even at the classical level.

In Sec.~\ref{sec:six} we turn to the final delocalization mechanism 
discussed in (2) above, which involves delocalization due to
a field becoming massless on the CMS. We consider 
a restriction of the two-field model, discussed in Sec.~\ref{sec:three} 
which, when 
perturbed by a term which breaks \ntwo to \none supersymmetry,
provides a simple example of this phenomenon. 

We collect some concluding remarks in Sec.~\ref{sec:seven}, and discuss in
particular the applicability of our results to marginal stability
of the $W$ boson, and also subtleties associated with extended BPS 
objects.

\section{Soliton Dynamics near the CMS}
\label{sec:two}

Before considering a specific model in detail, we first discuss some
simple but quite general constraints which are useful 
in providing a qualitative
guide to the dynamics appropriate to the near-CMS regime.

\subsection{Dominant interactions}
\label{domint}

Consider the dynamics of two primary BPS solitons with masses
$M_1$ and $M_2$ near a CMS curve for the composite BPS
soliton with mass $M_{1+2}$. From the CMS condition that the binding
energy vanishes, $M_{1+2}=M_1+M_2$, its clear that by going sufficiently
close to the CMS, the relevant energy scales -- kinetic and binding
energy -- can be made much smaller than the soliton masses. The system
is then non-relativistic, and 
the effective dynamics is supersymmetric
quantum mechanics on the space of collective coordinates of the
configuration. With spherically symmetric interactions, the relevant 
part of this system can be reduced to one dimensional SQM associated
with the relative separation $r$ between the primary solitons.

We can also deduce some generic features of the potential, in part from 
knowledge of the BPS mass spectrum. Firstly, its inconsistent for
the potential to be of attractive Coulomb-like form on the CMS itself.
This result follows straightforwardly from the incompatibility of the
BPS mass spectrum with the structure of the bound state energy levels 
associated with a Coulomb-like potential. Indeed the quantum
mechanical spectrum associated with the attractive $1/r$ potential
will exhibit towers of closely spaced bound states, only the lowest of
which can be BPS saturated. The Coulomb wavefunctions
$\psi \sim r^n e^{-r/n}$ lead to bound state levels
of  the form 
$\ep_n\propto -1/n^2$. In contrast, we know from the form 
of the BPS mass spectrum that on the CMS the lowest level 
in the tower must reach the continuum. Clearly the only way this can
happen is if the $1/r$ attractive  
interaction vanishes on the CMS.

In other words, if attractive Coulomb-like forces are generically
present,  there
must be a coefficient which we may identify as the distance to the
CMS,
\be
 V\left(r\right)_{\;\;\;r\to
   \infty}\!\!\!\!\!\!\!\!\!\!\!\!\longrightarrow \quad{\rm const} -
 (q^2- f)\frac{1}{4\pi\, r}
+\cdots, \label{lr}
\ee
where 
$q$ is used to denote the appropriate charge and $f$ is a
certain function of the moduli equal to $q^2$ on the CMS. 
The ellipsis denotes higher order terms in $1/r$.

A second constraint is the requirement that 
the potential admits a normalizable bound state arbitrarily
close to the CMS (inside the stability domain). This constraint
is dimension dependent. While in 1+1D and 2+1D an arbitrarily small 
attraction can result in such bound state, this is not the case in higher
dimensions.  
In 3+1D, in order to form a bound state the attraction must be strong enough,
$\int {\rm d}r \,r\, (-V(r))>\hbar^2/M$. In particular, for the 3+1D
dynamics of dyons in SU(3) SYM, as we will
see in Sec.~\ref{sec:five}, the bound state is due to a
Coulomb-like attraction at large 
distances \cite{ly,bhlms,tong,blly,bly,bl,n2quant}. Although according to
Eq.~(\ref{lr}) the effective Coulomb coupling
diminishes on approach to the CMS, the bound state does exist even 
for an arbitrarily small  coupling.

In conclusion, from the simple arguments above we deduce that 
close to the CMS the dynamics is nonrelativistic and the long range
component of the potential controlling the restructuring of the 
spectrum satisfies the following constraints. Firstly, on the CMS it 
either vanishes, or is repulsive. Secondly, the simplest way to 
form bound states in dimensions higher than 2+1D is for the potential
to have an attractive long range form off the CMS.

\subsection{The restructuring mechanism}
\label{sec:2.2}

To understand what happens to the spectrum in the near-CMS regime it will
be useful to present a simple model which exhibits the relevant
features. Specifically, we consider below the mechanism via which a
restructuring of the spectrum can occur.

Assume that the model under consideration contains a set of
parameters (to be denoted generically  as $\{\mu \}$), and
admits BPS solitons at a certain value $\{\mu_0 \}$.
The parameters $\{\mu \}$ can be moduli, or some parameters in the
action. The question is how can BPS solitons
disappear from the spectrum under continuous variations of $\{\mu \}$?
Generally speaking, we would expect that if the BPS state
exists at $\{\mu_0 \}$, it remains in the spectrum at least  in some
finite domain in the vicinity of $\{\mu_0 \}$.

The argument is based on the multiplicity of the corresponding
supermultiplet. Indeed, in the models to be considered below,
the number of states in the BPS multiplet is twice smaller than the
number of states in the non-BPS multiplet (this type of `shortening' is
typical). This means that
if a BPS state is to become non-BPS, a factor of two jump in the
number of states must occur.
Generally speaking, this will not happen
 under  continuous deformations of
$\{\mu \}$, unless from the very beginning we had
{\em two } BPS multiplets
which become degenerate at a certain point in the parameter space
and  combine together
to leave the BPS spectrum as a joint non-BPS multiplet.

We are more interested in another scenario -- when a BPS state
becomes non-BPS at a certain critical point $\{\mu_* \}$,
without the pre-arranged doubling of the type mentioned above.
Are we aware of any simple analogs of this phenomenon?

\begin{figure}[ht]
     \epsfxsize=7cm
  \centerline{   \epsfbox{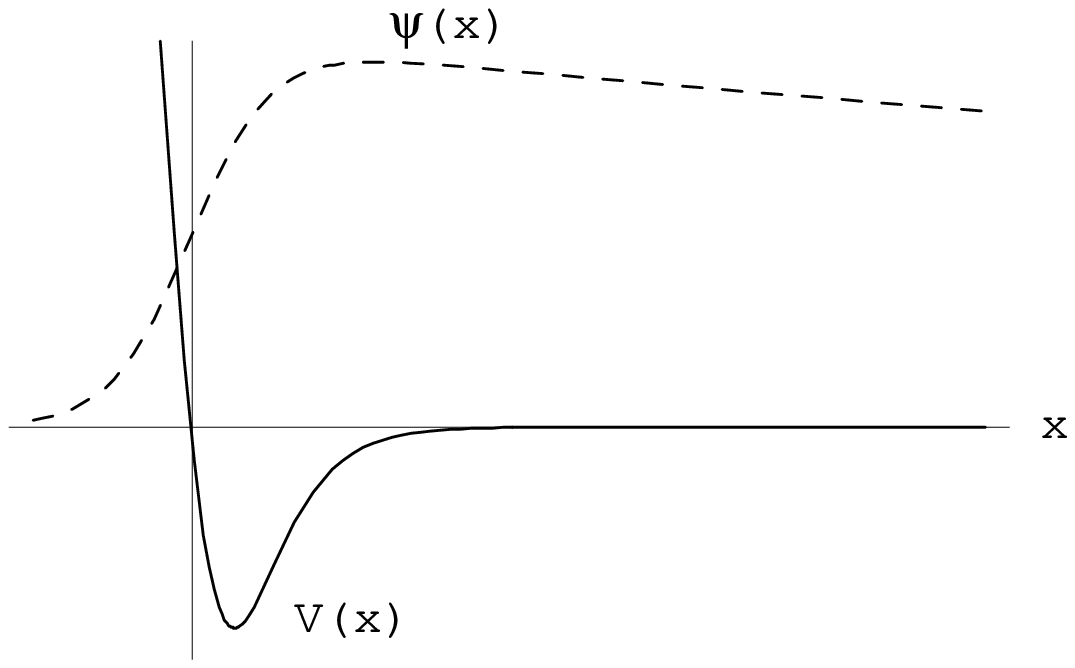}}
    \newcaption{The potential $V(x)$ in the
problem  (\ref{wqm}), (\ref{toysp}) (solid line)
and the corresponding ground state wave function (dashed line).
The parameter $\mu =0.98$. The units on the vertical and horizontal axes
are arbitrary.\label{fig:cmspotwf}}
\end{figure}

The answer is yes, a simple example has been known for a long time.
We will discuss it here for two reasons: firstly, it nicely illustrates
the generalities of the dynamical phenomenon discussed in the
preceding subsection; and secondly,
we will need to introduce the corresponding notation
later anyway. The example can be found in supersymmetric quantum
mechanics (SQM) with two supercharges introduced by Witten \cite{EWi}.
Consider a system (as motivated above) described by the Hamiltonian
\be
H = \frac{1}{2}\left[ p^2 + (W')^2 +  \sigma_3 W''\right]\,,
\label{wqm}
\ee
where $p= -i{\rm d}/{\rm d}x$, and $W$ is a function of $x$ with
the prime denoting differentiation by $x$. Moreover,
$\sigma_3$ is the third Pauli matrix corresponding to the fact that
$[\si_1,\si_2]$ forms an appropriate representation of the Grassmann
bilinear.
The function $W$ will be referred to as the SQM superpotential.
Two  conserved (real) supercharges are
\be
Q_1= \frac{1}{\sqrt{2}}\left(p\,\sigma_1  +  W'\, \sigma_2\right) \,,\qquad
Q_2= \frac{1}{\sqrt{2}}\left(p\, \sigma_2  -  W' \, \sigma_1\right)\,.
\ee
They form the following superalgebra,
\be
\left(Q_1\right)^2=\left(Q_2\right)^2=H\;,\qquad
\left\{Q_1,Q_2\right\}=0\;.
\label{sqmalg}
\ee
If $W'$ has an odd number of zeros then the ground state of the system
(\ref{wqm}) is supersymmetric (i.e. the supercharges annihilate it)
and unique. This is the analog of the BPS soliton. If  $W'$ has no zeros 
or even number of zeros,
the ground state is doubly degenerate and is not annihilated by the
supercharges. The ground states in this case are analogs of non-BPS
solitons. The unique versus doubly degenerate ground  state
in the problem (\ref{wqm}) imitates ``multiplet
shortening''. The transition from the first case to the second
under continuous deformations of parameters is easy to visualize.
Indeed, let us assume, for definiteness,
that
\be
W(x) = 
 \ln \cosh  x - \mu x
\,, \quad W' =  
\tanh x - \mu \;.
\label{toysp}
\ee
At $\mu = 0$ the derivative of the superpotential vanishes at the
origin.
As $\mu$ grows (remaining positive), the point where $W'$ vanishes
shifts to the right, towards large positive  values of
$x$. The
ground state wave function is supersymmetric and unique,
\be
\Psi_0 = {\rm e}^{- W(x)}\, |\downarrow \rangle\, = \frac{e^{\mu
x}}{\cosh x} | \downarrow\rangle\,.
\label{wfzero}
\ee
As one approaches $\mu_* = 1$ from below this wave function becomes
flatter on the right semi-axis; representing a swelling of the
bound state in coordinate space. The corresponding scalar potential
$$
V(x) = \frac 1 2\left[ (W')^2 -  W''\right]
\label{scalpot}
$$
at $\mu = 0.98$ and the
ground state wave function are depicted in Fig.~\ref{fig:cmspotwf}.

The point $\mu = 1$ is  critical. At $\mu > 1$
the wave function (\ref{wfzero}) at $E=0$ becomes non-normalizable,
and the true ground state, coinciding  with the continuum threshold,
spectrum is doubly degenerate.
The transition from one regime to another occurs through
delocalization in that the zero of $W'$, the equilibrium
point $x_0$, escapes to infinity.
Note that dynamically the SQM problem under consideration is similar
to that of deuterium. The potential well in  Fig.~\ref{fig:cmspotwf} is
at $x < 1$, but the tail of the wave function stretches very far to the
right due to the fact that the $E=0$ level is very close to the
continuum spectrum.

We can make this somewhat more precise by introducing a ``classicality
parameter'' $\xi$ defined as
\be
 \xi \equiv \left.\frac{W''''(x)}{(W''(x))^2}\right|_{x=x_0},
  \label{xidef}
\ee
where $x_0$ is the classical minimum of the potential: $W'(x_0)=0$.
The parameter $\xi$ may be interpreted as measuring the quantum correction 
to the curvature  of the
potential at the classical equilibrium point. i.e. the system
is essentially classical if $\xi\ll 1$, while it is highly quantum
if $\xi\gg 1$.

In the current example, we find that as we approach the critical point,
\be
 \xi \sim \frac{1}{2\,(1-\mu)}+\cdots,
\ee
and so the system indeed becomes highly quantum in this regime.
In fact this feature is quite generic for short range potentials
and may be viewed as an artifact of the remnant supersymmetry.
Specifically, since the mass term $V_F\sim W''(x)$ is
linear in the superpotential, while the bosonic potential is
quadratic, $V_B\sim (W'(x))^2$, for short range
interactions the fermionic $W''$ term in the superpotential (\ref{scalpot})
will dominate for large
separations. This is despite the fact that the fermionic term
is a quantum effect (in field theory it corresponds to the 1-loop
correction to the effective potential through integrating out the
fermions).  Thus, although the system becomes more and more weakly
bound, in the CMS region the system enters a highly quantum regime
where the classical minimum of the bosonic potential need not be
relevant.
Below we will see that exactly the same phenomenon occurs for 
BPS solitons near the CMS in a 1+1D Wess-Zumino model.

\subsection{Embedding of SQM within the field theory superalgebra}
\label{sec:2.3}

To establish a link between the field-theoretical description of
solitons on the one hand and the supersymmetric quantum mechanics
of two nonrelativistic primary states on the other, we now consider
the manner in which the quantum mechanical supercharges emerge from the
full field-theoretical superalgebra. The fact that supersymmetry is
realized linearly in the two soliton sector may be reinterpreted as
the existence of a straightforward embedding of the SQM supercharges.
Moreover, near the CMS the system becomes essentially nonrelativistic
and we need keep only the leading term in an expansion in velocities.

Although the arguments apply more generally, we consider for
definiteness the realization of \ntwo supersymmetry in two dimensions
in the two soliton sector. Recall that the algebra contains
four supercharges $Q_\alpha$,
$Q^\dagger_\alpha$ ($\alpha=1,2$) and has the form \cite{Ken,Vafa}
\begin{equation}
\{Q_\alpha,Q^\dagger_\beta\}
=2\,(\gamma^\mu\gamma^0)_{\alpha\beta}\,P_\mu\,,\quad \!\!
\{Q_\alpha,Q_\beta\}=2\, i
\,(\gamma^5\gamma^0)_{\alpha\beta}\,\bar{\cal Z}\,,\quad\!\!
\{Q^\dagger_\alpha,Q^\dagger_\beta\}=2\, i
\,(\gamma^5\gamma^0)_{\alpha\beta}\,{\cal Z}\,,
\label{alg0}
\end{equation}
where $P_\mu=(P_0,P_1)$ are the energy-momentum operators and
${\cal Z}$ is a complex central charge.
We use the Majorana  basis for 2$\times$2 $\gamma$-matrices,
\begin{equation}
\gamma^0\!=\!\sigma_2\;,\quad
\gamma^1\!=\!{ i}\,\sigma_3\;,\quad
\gamma^5=\gamma^0\gamma^1=-\sigma_1\;.
\end{equation}

Modulo addition of the central charge, the algebra (\ref{alg0}) can be viewed
as a dimensional reduction of the \none algebra in four dimensions.
The SO(3,1) Lorentz symmetry in 3+1 dimensions
reduces in 1+1  to the product SO(1,1)$\times$U(1)$_R$ where SO(1,1)
is the Lorentz boost in 1+1 and U(1)$_R$ is a
global symmetry associated with the fermion charge. More precisely,
the Lorentz boost with parameter $\beta$ acts on the supercharges
$Q_\alpha$ as follows
\begin{equation}
Q_\alpha \to \left[\exp\left(\frac{1}{2}\,\beta\,
\gamma^5\right)\right]_{\alpha\beta}\, Q_\beta\;,
\label{boost1}
\end{equation}
while the U(1)$_R$ transformation 
with parameter $\eta$ is 
\begin{equation}
 Q_\alpha \to \left[\exp\left(\frac{i}{2}\,\eta\,
\gamma^5\right)\right]_{\alpha\beta}\, Q_\beta\;.
\label{ferm1}
\end{equation}
Notice that the U(1)$_R$ transformations can be viewed as a
complexification of the Lorentz boost (\ref{boost1}).

Its now convenient to introduce the Majorana  supercharges
$Q_\alpha^i$,
($i=1,2\,$;
$(Q_\alpha^i)^\dagger=Q_\alpha^i$) via the relation
\begin{equation}
Q_\alpha=\frac{{\rm e}^{-i\,\alpha/2}}{\sqrt{2}}\left(Q^1_\alpha + i\,
Q^2_\alpha\right)\;,
\label{realch}
\end{equation}
where the phase factor ${\rm e}^{-i\,\alpha/2}$ contains an arbitrary
parameter $\alpha$, which we will fix momentarily.
In terms of  $Q_\alpha^i$ the algebra (\ref{alg0}) has the form:
\begin{equation}
\{Q_\alpha^i,Q_\beta^j\}=2\,\delta^{ij}(\gamma^\mu\gamma^0)_{\alpha\beta
}\,
P_\mu+ 2\, i\,(\gamma^5\gamma^0)_{\alpha\beta}\,{\cal Z}^{ij}\;,
\label{alg2}
\end{equation}
where the $2\times 2$ real matrix  of central charges ${\cal Z}^{ij}$ is
symmetric and traceless. It is related to the original complex ${\cal
Z}$ as follows,
\begin{equation}
{\cal Z}{\rm
e}^{-i\,\alpha}={\cal Z}^{11}-i\, {\cal Z}^{12}\;.
\end{equation}

To consider representations of the algebra we use a Lorentz boost in
1+1 to put the system in the rest frame where $P_1\to 0$ and $P_0\to
M=\sqrt{P_\mu P^\mu}$. Moreover, we can always choose the basis in 
U(1)$_R$ to put ${\cal Z}^{ij}$ in the form ${\cal Z}^{ij}=|{\cal Z}|
\,\tau_3^{ij}$.
This amounts to fixing the phase $\alpha$ to be equal to the phase of the
central charge, ${\cal Z}=|{\cal Z}|\,{\rm e}^{i\,\alpha}$.
Then the algebra (\ref{alg2}) takes the following component form,
\begin{equation}
\left(Q_1^1\right)^2=\left(Q_2^2\right)^2=M+|{\cal Z}|\;,\qquad
\left(Q_2^1\right)^2=\left(Q_1^2\right)^2=M-|{\cal Z}|\;,
\label{alg3}
\end{equation}
with all other anticommutators vanishing, so that
the algebra splits into two independent subalgebras.

>From (\ref{alg3}) we see that $|{\cal Z}|$ is a lower
bound for the mass, $M\ge |{\cal Z}|$. When $M> |{\cal Z}|$ the
irreducible representation  has dimension four -- two bosonic and two
fermionic states. The BPS states  saturate the lower bound, $M_{\rm
BPS}= |{\cal Z}|$, and in this case the second subalgebra becomes trivial
and the representation is two-dimensional -- one bosonic and one
fermionic state \cite{Ken}.

How do all of these generalities help us with the problem of
constructing the SQM near the
CMS? In the vicinity of the CMS the difference $M-|{\cal Z}|$ is small
as compared to
$|{\cal Z}|$ and can be identified  with the nonrelativistic
Hamiltonian,\footnote{Note that we view $M$ as a Hilbert space 
operator.}
\begin{equation}
H_{\rm
SQM}=M-|{\cal Z}|\;.
\label{hsqm1}
\end{equation}
The second subalgebra in Eq.~(\ref{alg3}) with supercharges $Q_1^1$ and
$Q_2^2$ then coincides with that of the standard SQM, see
Eq.~(\ref{sqmalg}). In the first subalgebra the operator $M+|{\cal Z}|$
can be substituted by $2\,|{\cal Z}|$ up to relativistic corrections. 
Consequently, the first subalgebra just leads to a generic
multiplet structure (in this case just duplication) 
for every state found in the SQM.

In Sec.~\ref{sec:four} we will find all the supercharges in the 1+1 example
as explicit functions of the moduli from field-theoretic  solutions
for two solitons, $u$ and $v$. Near the CMS, where their relative 
motion is nonrelativistic, the result can
be compared with the quantum mechanical realization of the superalgebra
(\ref{alg3}). For $ H_{\rm SQM}=M-|{\cal Z}|$ we take the expression which
generalizes Eq.~(\ref{wqm}) to include a mass parameter,
\begin{equation}
H_{\rm SQM} = \frac{1}{2M_r}\left[ p^2 + (W')^2 + 
\sigma_3 \,W''\right]\,,
\label{hsqm}
\end{equation}
where the superpotential $W$ depends on the separation $s=z_u-z_v$,
the conjugate momentum $p=-i{\rm d}/{\rm d}s$, and $M_r$ is the reduced
mass,
\begin{equation}
 M_r = \frac{M_u\,M_v}{M_u+M_v}\;.
\label{reduce}
\end{equation}
 Then a realization of the superalgebra can be chosen in the 
form ($\si_i$ and $\ta_i$ are two sets of Pauli matrices):
\begin{eqnarray}
&& Q_1^1=\sqrt{2\,|{\cal Z}|}\,\tau_1\otimes \sigma_3\;,\qquad \qquad
\qquad
 ~~Q_2^2=\sqrt{2\,|{\cal Z}|}\,\tau_2\otimes \sigma_3\;,
\nonumber\\[2mm]
&& Q_2^1=I\otimes\frac{1}{\sqrt{2\,M_r}} \left[p\,
\sigma_1+ W'(s)\,\sigma_2\right]\;,\quad
 Q_1^2=I\otimes\frac{1}{\sqrt{2\,
M_r}}\left[p\,\sigma_2-W'(s)\,\sigma_1\right]\;.
\label{reali}
\end{eqnarray}

The realization (\ref{reali}) explicitly indicates a
factorization of both the bosonic and fermionic degrees of freedom
associated with the center of mass of the system.  We can also include
dependence on the total spatial momentum $P_1$  through a Lorentz
boost (\ref{boost1}) with  $\tanh\beta=P_1/\sqrt{M^2+P_1^2}$.

A couple of comments are now in order. Firstly, its clear from this
construction that the SQM can only be realized linearly in BPS
sectors with a non-vanishing central charge. Otherwise, one has 
$Q=\sqrt{M} \, \psi$  (with $\psi$ a
fermionic operator) in the nonrelativistic limit, implying a 
nonlinear realization. Secondly, we note that
the expressions for $Q^1_1$ and $Q^2_2$ in the first line of  
Eq.~(\ref{reali}) represent the leading
terms in the nonrelativistic $v/c$ expansion. It is not 
difficult to include higher order terms in this expansion
as follows,
\begin{eqnarray}
&&Q_1^1=\tau_1\otimes \sigma_3\!\left[2\,|{\cal Z}|+\frac{p^2+(W')^2+
 \sigma_3 W''}{2M_r}\right]^{1/2}\;,\nonumber\\[1mm]
&& Q_2^2=\tau_2\otimes \sigma_3\!\left[2\,|{\cal Z}|+\frac{p^2+(W')^2+
 \sigma_3 W''}{2M_r}\right]^{1/2}
\;.
\label{relcor}
\end{eqnarray}
where the square root is to be understood as an 
expansion in $1/|{\cal Z}|$.

In concluding this section, we note that within the context
of the present \mbox{\ntwo} system one can formulate a general statement: 
given the subalgebra (\ref{sqmalg}) with two supercharges it is 
always possible to elevate it to a superalgebra with
four supercharges and a central charge ${\cal Z}$ by adding the 
two additional supercharges (\ref{relcor}).

\section{An ${\cal N}=2$ WZ Model in Two Dimensions}
\label{sec:three}

\subsection{Introducing the model}
\label{sec:3.1}

With the aim of concretely illustrating the general arguments of the
previous section, we now consider a specific model. A suitable
example exists in two dimensions, obtained by
dimensional reduction of a four-dimensional
Wess-Zumino model with two chiral superfields. The latter is
the deformation of a model  considered previously in
Ref.~\cite{Shifman:1998wg}. The superpotential is
\begin{equation}
{\cal W}(\Phi,X)=\frac{m^2}{\lambda}\,\Phi-\frac{\lambda}{3}\,\Phi^3
-\lambda\,\Phi\,X^2 +
\mu\,m\,X^2 +\frac{m^2}{\lambda}\,\nu \,X\,,
\label{wphix}
\end{equation}
where $\Phi$ and $X$ are two chiral superfields, $m$ is a mass
parameter,
$\lambda$ is the coupling constant, while $\mu$ and $\nu$
are deformation parameters. By an appropriate phase rotation of fields
and the superpotential one can always make $m$ and $\lambda$ real and
positive.
The parameters  $\mu$ and $\nu$ are in general complex,
\begin{equation}
\mu \equiv \mu_1 + i \mu_2\,,\quad \nu \equiv \nu_1 + i \nu_2\, .
\end{equation}
The four real dimensionless parameters $\mu_1$, $\mu_2$, $\nu_1$ and
$\nu_2$
will form our parameter space $\{\mu\}$. For technical reasons the
parameter
$\mu \equiv \mu_1 + i \mu_2$ will be assumed to be small in what
follows,
$\mu_{1,2}\ll 1$. Furthermore, we will consistently work in
the approximation in which the superpotential is linear
in $\mu$; this corresponds to terms of $O(\mu^2)$ in the scalar
potential.
This limitation is not a matter of principle but, rather, for
technical convenience. In the limit of small $\mu$ we can obtain all
formulae in closed form.
We will also take the coupling constant $\lambda$
to be small, $\lambda/m \ll 1$, so that a quasiclassical treatment is
applicable (except on
some exceptional submanifolds in the parameter space).

As a two-dimensional model, this theory has extended
\ntwo supersymmetry, and exhibits solitonic 
kinks interpolating between the distinct vacua. In two dimensions they
are particles (in four dimensions they would be domain walls).
The dimensionality of the BPS supermultiplet is two, while that of
the non-BPS supermultiplet is four.

Substituting
\begin{equation}
\Phi=\frac{m}{2\lambda}\,(U+V)\,,\qquad X=\frac{m}{2\lambda}\,(U-V)\,,
\end{equation}
we arrive at the following action,
\begin{equation}
S=\frac{m^2}{2\lambda^2}\left\{\frac 1 4 \int{\rm d}^2 x \,{\rm
d}^4\theta
\,({\bar U}U+{\bar V}V) +\left(\frac m 2 \int{\rm d}^2 x \,{\rm
d}^2\theta\, {\cal
W}(U,V)+{\rm h.c.}\right)\right\}\,,
\label{uvact}
\end{equation}
where the dimensionless superpotential ${\cal W}$ is
\begin{equation}
{\cal W}(U,V)= U-\frac 1 3\, U^3+ V-\frac 1 3\, V^3
+\frac{\mu}{2}\,(U-V)^2
+\nu\,(U-V)\;.
\end{equation}
The vacua of the model are defined by $\partial{\cal W}/\partial u=0$,
$\partial{\cal W}/\partial v=0$,
\begin{eqnarray}
&&1+\nu -u^2 +\mu\,(u-v)=0\;, \nonumber\\[0.1 cm]
&&1-\nu -v^2 -\mu\,(u-v)=0\;.
\end{eqnarray}

\begin{figure}[ht]
\epsfxsize=11cm
 \centerline{%
   \epsfbox{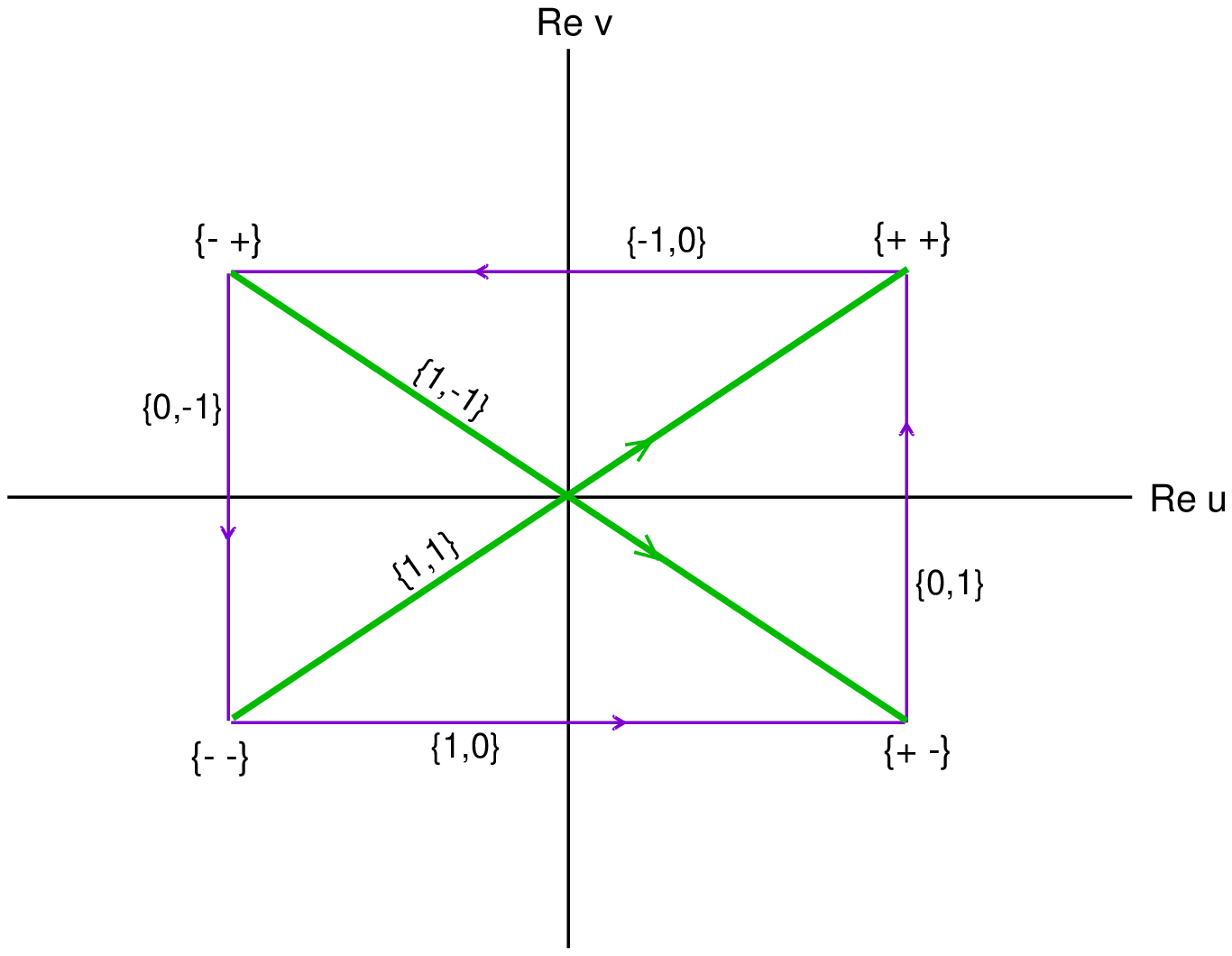}%
         }
 \newcaption{Structure of vacua and solitons in the ${\rm Re}\, u$,
${\rm Re}\, v$
plane for real $\nu$ and $\mu$ .\label{fig:vac}}
\end{figure}

For real $\mu$ and $\nu$  the solutions to these equations define four
different
vacua with real values of the fields and real values of the
superpotential
${\cal W}(u,v)$. The vacuum structure is illustrated in
Fig.~\ref{fig:vac} for small $\mu$.
One of these vacua (denoted as $\{++\}$ in Fig.~\ref{fig:vac} )
corresponds to a maximum of the real function
${\cal W}(u,v)$ on the real section of the variables $u$ and $v$, the
other vacuum (denoted as $\{--\}$)
corresponds to a minimum of ${\cal W}(u,v)$, and the remaining two
vacua ($\{+-\}$ and $\{-+\}$) are saddle points. 

In this situation there
exists \cite{Shifman:1998wg} a continuous family of real BPS solitons, i.e. of
solutions to the real BPS equations
\beq
\frac 1 m \,{{\rm d} \over {\rm d} z} \,u = {\partial {\cal W} \over
\partial u}\;,
\qquad
\frac 1 m \,{{\rm d} \over {\rm d} z}\,v = {\partial {\cal W} \over
\partial v}\;,
\label{bpsreal}
\eeq
interpolating between the $\{--\}$ vacuum with ${\cal W}_{\rm min}$ at
$z=-\infty$ and the $\{++\}$ vacuum with ${\cal W}_{\rm max}$ at
$z=\infty$. All
these solitons are  degenerate in mass: $M={\cal W}_{\rm max}-
{\cal W}_{\rm min}$, and can be viewed as a superposition of non-interacting
primary solitons: one going from the vacuum with ${\cal W}_{\rm min}$ to
one of the saddle points, and the other soliton going from the saddle point
to the vacuum with ${\cal W}_{\rm max}$. The parameter labeling the
solutions in this family can be interpreted in terms of the distance
between the basic solitons, and thus the degeneracy in energy implies that
there is no interaction between the basic solitons at real $\mu$ and $\nu$, at
least for some finite range of the these parameters.
The decoupling of the dynamics of the primary solitons at $\mu=0$ is
trivial, as the superfields $U$ and $V$ are also decoupled within the 
underlying field theory. However, at $\mu_1\neq0$, there is no such
decoupling within the field theory but, nevertheless, the primary solitons 
do not interact at rest (provided $\mu$ and $\nu$ are real). This is a
manifestation of the nontrivial ``no-force'' condition for BPS states.

\subsection{Decoupled solitons, $\mu = 0$ case}

At $\mu=0$ the model is extremely simple: the fields $U$  and $V$ are
not coupled. Their vevs are
\begin{equation}
u=\pm\,\nu_+\,,\qquad v=\pm\,\nu_-\;,
\end{equation}
where we introduce the notation
\begin{equation}
\nu_{\pm}=\sqrt{1\pm\nu}\,.
\label{nupm}
\end{equation}
The masses  of the BPS solitons are given by
\begin{equation}
M_{n_u,n_v}={4 \over 3} \,\frac{m^3}{\lambda^2} \left| \,n_u \nu_+^3+n_v
\nu_-^3\right|\,,
\label{bpsmass}
\end{equation}
where the topological charges are $n_{u,v}=0,\pm1$ (see
Fig.~\ref{fig:vac}).

\begin{figure}[ht]
     \epsfxsize=7cm
     \centerline{     \epsfbox{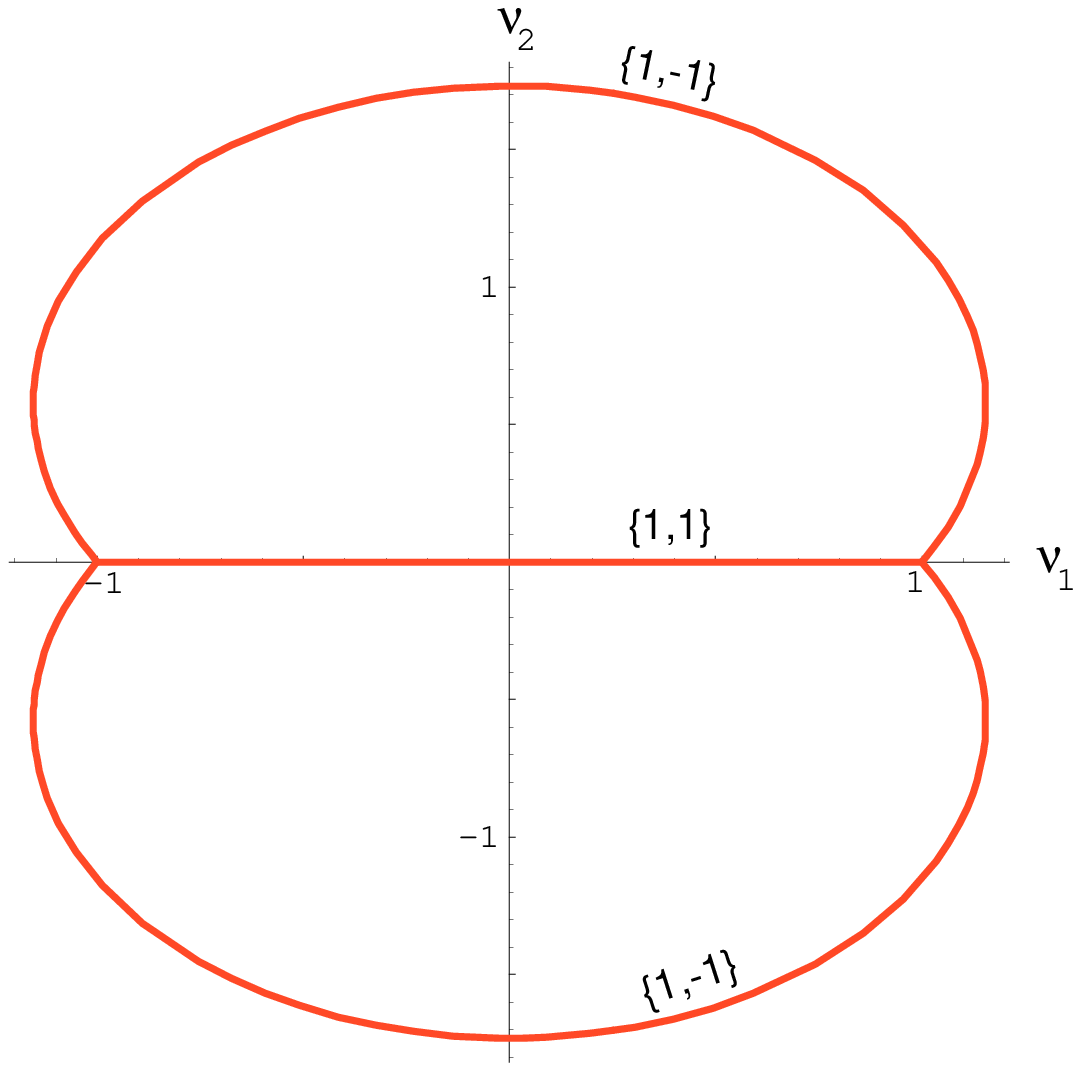}}
    \newcaption{The curve of marginal stability in the complex plane of
$\nu$. \label{fig:cmsnunu}}
\end{figure}

However, as noted in \cite{{Shifman:1998wg}},  not all combinations of charges
are realized. For a generic value of the complex parameter $\nu$
only the $\{1,0\}$ and  $\{0,1\}$ solitons and their antiparticles
exist as BPS
states. To have a BPS state with both $n_u$ and $n_v$ nonvanishing, one needs
to align in the complex plane the two 
terms, $\nu_+^3$ and $\nu_-^3$, contributing
to the mass in Eq.~(\ref{bpsmass}). The relevant conditions are
\begin{equation}
{\rm Im}\,\left(\frac{\nu_-}{\nu_+}\right)^{3}=0\;,
\qquad \frac{n_v}{n_u}\,{\rm
Re}\,\left(\frac{\nu_-}{\nu_+}\right)^{3}>0\;.
\label{align}
\end{equation}
These conditions define a curve  in the complex
$\nu$ plane presented in Fig.~\ref{fig:cmsnunu}.
This curve is the curve of marginal stability for the model. In the case under
consideration, with no interaction, the CMS coincides with stability
domains for composite solitons, they only  exist on this curve.

The curve in Fig.~\ref{fig:cmsnunu} consists of three parts which can be
parametrized as
\begin{equation}
\nu=\tanh \sigma\,,\qquad {\rm Im} \,\sigma=0,\,\pm \frac \pi 3\,.
\label{cmszero}
\end{equation}
The part sitting on the real axis between $\nu=\pm 1$ (corresponding to
${\rm Im}\,\sigma=0$) is the stability domain for the $\{1,1\}$ composite
solitons (and their antiparticles). The other two parts,   
${\rm Im}\,\sigma=\pm \,\pi/3$, give
the stability domain for the  $\{1,-1\}$ and $\{-1,1\}$ solitons.

 The bifurcations at $\nu=\pm 1$ are due to the vanishing of   
the mass of one of the 
primary solitons at these points.  It is explained by the degeneracy of
vacua  at these values -- instead of four vacua only two remain at
$\nu=\pm 1$ (strictly speaking there are still four, but they coalesce
in pairs). These are simple analogs of the Argyres-Douglas points
\cite{ad} in  gauge
theories.

\subsection{Stabilization by  $\mu$}

The model at $\mu=0$ is a very degenerate case. Indeed, the extra
$\{\pm\,1,\pm\,1\}$ states exist {\em only} on the CMS and are nothing but
systems of two noninteracting $\{\pm\,1,0\}$ and$\{0,\pm1\}$ solitons.  The
relative separation between the primary solitons 
is an extra classical modulus, on quantum
level the $\{\pm\,1,\pm\,1\}$ solitons are not localized states.
As we will show, the introduction of a
nonvanishing ${\rm Im}\mu=\mu_2$ expands the domain of stability for the
extra BPS state which then occupies a finite area near the original curve.
Thus, setting $\mu_2$ nonzero leads to  an attraction of the primary solitons.

Using $\mu$ as a perturbation parameter we find the vevs and values of the
superpotential ${\cal W}$ for the four vacua to first order in
$\mu$.
\begin{eqnarray}
&&\{++\}: ~~~~u=\nu_++\frac \mu
2\left(1-\frac{\nu_-}{\nu_+}\right),\quad v=\nu_-+\frac
\mu 2\left(1-\frac{\nu_+}{\nu_-}\right),\nonumber\\[0.1cm]
&&~~~~~~~~~~~~~~{\cal W}_{++}=\frac 2 3\,\nu_+^{3}+\frac 2
3\,\nu_-^{3}+\mu\left(1-\nu_-\nu_+\right);\nonumber\\[0.2cm]
&&\{+-\}:  ~~~~u=\nu_++\frac \mu
2\left(1+\frac{\nu_-}{\nu_+}\right),\quad v=-\nu_-+\frac
\mu
2\left(1+\frac{\nu_+}{\nu_-}\right),\nonumber\\[0.1cm]
&&~~~~~~~~~~~~~~{\cal W}_{+-}=\frac 2 3\,\nu_+^{3}-\frac 2
3\,\nu_-^{3}+\mu\left(1+\nu_+\nu_-\right);\nonumber\\[0.2cm]
&&\{-+\}:  ~~~~u=-\nu_++\frac \mu
2\left(1+\frac{\nu_-}{\nu_+}\right),\quad v=\nu_-+\frac
\mu
2\left(1+\frac{\nu_+}{\nu_-}\right),\nonumber\\[0.1cm]
&&~~~~~~~~~~~~~~{\cal W}_{-+}=-\frac 2 3\,\nu_+^{3}+\frac 2
3\,\nu_-^{3}+\mu\left(1+\nu_+\nu_-\right);\nonumber\\[0.2cm]
&&\{--\}: ~~~~u=-\nu_++\frac \mu
2\left(1-\frac{\nu_-}{\nu_+}\right),\quad v=-\nu_-+\frac
\mu
2\left(1-\frac{\nu_+}{\nu_-}\right),\nonumber\\[0.1cm]
&&~~~~~~~~~~~~~~{\cal W}_{--}=-\frac 2 3\,\nu_+^{3}-\frac 2
3\,\nu_-^{3}+\mu\left(1-\nu_+\nu_-\right).
\label{vacua}
\end{eqnarray}
The BPS masses are given by $|{\cal W}_{ij}-{\cal W}_{i'j'}|$ and the alignment
conditions which define the CMS to first order in $\mu$ become (c.f.
Eq.~(\ref{align})),
\begin{equation}
{\rm
Im}\,\left(\frac{\nu_-^2-\mu\,\nu_+}{\nu_+^2+\mu\,\nu_-}\right
)^{3/2}=0\;,
\qquad
{\rm
Im}\,\left(\frac{\nu_-^2+\mu\,\nu_+}{\nu_+^2-\mu\,\nu_-}\right
)^{3/2}=0\;~,
\label{align1}
\end{equation}
where the conditions clearly differ only by a choice of the branch of
the square root in the terms linear in $\mu$.
Analytical expressions for the CMS are simpler
in terms of the complex parameter of $\sigma$ (related to $\nu$ by
Eq.~(\ref{cmszero})). In the  complex $\si$-plane the CMS is given by
the curves
\begin{eqnarray}
\sigma_2& = & \pm \,\mu_2\, \cosh \frac
{3\sigma_1}{2}\,\cosh^{1/2}\sigma_1\;,\nonumber\\[0.1cm]
\sigma_2& = &~\frac \pi 3 \pm \,\sinh \frac
{3\sigma_1}{2}\,{\rm Re}\left[\mu\, \cosh^{1/2}\left(\sigma_1+i\,\frac
\pi
3\right)\right]\;,\nonumber\\[0.1cm]
\sigma_2& =& -\frac \pi 3 \pm \,\sinh \frac
{3\sigma_1}{2}\,{\rm Re}\left[\mu\, \cosh^{1/2}\left(\sigma_1-i\,\frac
\pi
3\right)\right]\;,
\end{eqnarray}
where the indices 1 and 2 refer to the real and imaginary parts,
$\sigma=\sigma_1+i\sigma_2$.

\begin{figure}[ht]
     \epsfxsize=7cm
     \centerline{\epsfbox{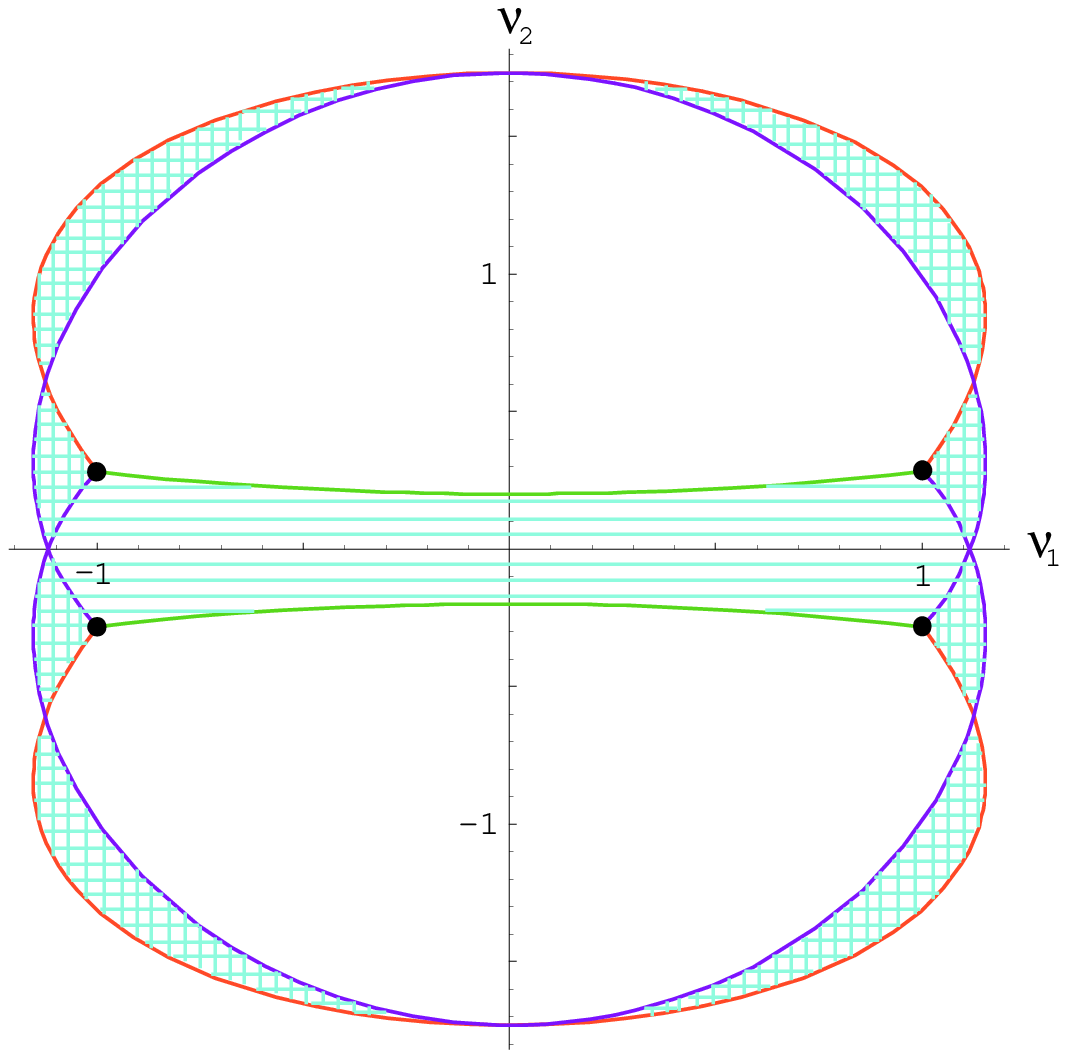}}
    \newcaption{The domains of stability for the composite BPS states
(shown for
$\mu_2=0.2$). The hatched region along the real axis is the stability
domain for the $\{1,1\}$ solitons and its antiparticles, in the cross
hatched one the $\{1,-1\}$ solitons and its antiparticles are
stable.\label{fig:cmsf2}}
\end{figure}

The  curves of marginal stability in the $\nu$ plane are presented in
Fig.~\ref{fig:cmsf2}. They form the boundaries of the stability domains
for the
composite BPS states marked in the figure.
Figure~\ref{fig:cmsf2} exemplifies different metamorphoses of the composite
BPS solitons on the CMS: crossing some boundaries leads to disappearance of
the BPS state from the spectrum, on others the original BPS state
disappears but a new one appears.

The figure also shows exceptional points on the CMS,  where two
stability subdomains of  the same BPS soliton touch each other. We shall
address a dynamical scenario at such points in Sec.~\ref{sec:4.3}.
 Note also four points of bifurcation (the Argyres-Douglas points)
where a pair of the vacuum states collide.

\subsection{A loosely bound composite BPS state}

In this subsection we will find a solution to the BPS equations for the
composite $\{1,1\}$ soliton. The construction explicitly demonstrates
that in the
vicinity of  the CMS this soliton is a loosely bound state of the primary
constituents. For definiteness we choose the region near the 
real $\nu$-axis and
the $\{--\}\to \{++\}$ transition. The BPS equations have the form
\begin{eqnarray}
\frac{1}{m}\,\frac{{\rm d}\, u}{{\rm d}z}= {\rm
e}^{i\alpha}\left[1+\nu^* - (u^*)^2
+\mu^*(u^*-v^*)\right]
\nonumber\\[0.2cm]
\frac{1}{m}\,\frac{{\rm d}\, v}{{\rm d}z}= {\rm
e}^{i\alpha}\left[1-\nu^* - (v^*)^2
-\mu^*(u^*-v^*)\right]
\label{uveqs}
\end{eqnarray}
where
\begin{equation}
{\rm e}^{i\alpha}=\sqrt{\frac{{\cal W}_{++}-{\cal W}_{--}}{{\cal
W}_{++}^*-{\cal
W}_{--}^*}}=\frac{\nu_+^{3}+\nu_-^{3}}{\left|\nu_+^{3}+
\nu_-^{3}\right|}\;.
\end{equation}
We will use perturbation theory in $\mu$.
The part of the CMS chosen for consideration at zeroth order in
$\mu$ corresponds to real $\nu$: $-1 < \nu_1 < 1$,
$\nu_2=0$. Then, at this order, $\alpha=0$ and the solution for $u$ and
$v$ reads 
\begin{equation}
u^{(0)}=\nu_{1+}\tanh\left[\nu_{1+}\,m\,(z-z_u)\right]\;,\quad
v^{(0)}=\nu_{1-}\tanh\left[\nu_{1-}\,m\,(z-z_v)\right]\;,
\label{lead}
\end{equation}
where $\nu_{1\pm}=\sqrt{\nu_1\pm 1}$ (see Eq.~(\ref{nupm})) and
the parameters $z_u$ and $z_v$ are arbitrary and denote the
positions of the centers of
the $u$- and $v$-solitons.

At first order in $\mu$, the soliton solutions become complex.
With an expansion about the leading order solutions $u^{(0)}$, $v^{(0)}$
of the form
\begin{equation}
u=u^{(0)}+(u_1+iu_2)+\cdots\;,\qquad v=v^{(0)}+(v_1+iv_2)+\cdots,
\end{equation}
the equations
(\ref{uveqs}) lead to
\begin{eqnarray}
&&\frac{1}{m}\,\frac{{\rm d}}{{\rm d}z}\,u_1=
-2u^{(0)}\, u_1
 +\mu_1\left(u^{(0)}-v^{(0)}\right)\;,
\nonumber\\[0.1cm]
&&\frac{1}{m}\,\frac{{\rm d}}{{\rm d}z}\,v_1=
-2v^{(0)}\,v_1
-\mu_1\left(u^{(0)}-v^{(0)}\right)\;,
\nonumber\\[0.1cm]
&&\frac{1}{m}\,\frac{{\rm d} }{{\rm d}z}\,u_2=2u^{(0)}u_2 -\nu_2
+\alpha\left[\nu_{1+}^2-(u^{(0)})^2\right]
-\mu_2\left[u^{(0)}-v^{(0)}\right]\;,
\nonumber\\[0.1cm]
&&\frac{1}{m}\,\frac{{\rm d}}{{\rm d}z}\, v_2=2v^{(0)}v_2 +\nu_2
+\alpha\left[\nu_{1-}^2-(v^{(0)})^2\right]
+\mu_2\left[u^{(0)}-v^{(0)}\right]\;,
\label{bps-mu}
\end{eqnarray}
where
\begin{equation}
\alpha=\frac 3 2 \,
\nu_2\,\frac{\nu_{1+}-\nu_{1-}}{\nu_{1+}^{3}+\nu_{1-}^
{3}}\;.
\label{alpha}
\end{equation}
Let us consider the equation for $u_2$. The function
$\cosh^2[\nu_{1+}\,m(z-z_u)]$ is the solution of the  homogeneous
part of
this equation, and the full solution is
\begin{eqnarray}
u_2(z)&=&\cosh^2\left[\nu_{1+}\,m\,(z-z_u)\right]\times
\nonumber\\[0.1cm]
&&m\int_{-\infty}^z\!\!{\rm d}x
\,\frac{-\nu_2 +\alpha\left[\nu_{1+}^2-(u^{(0)}(x))^2\right]
-\mu_2\left[u^{(0)}(x)-v^{(0)}(x)\right]}
{\cosh^2\left[\nu_{1+}\,m\,(x-z_u)\right]}
\end{eqnarray}
As $z\to -\infty$ the solution satisfies the boundary condition
\begin{equation}
\lim_{z\to
-\infty}u_2(z)=-\frac{\nu_2}{2\nu_+}+\frac{\mu_2}{2}\left(
1-\frac{\nu_{1-}}{\nu_{1+}}\right)
\end{equation}
consistent with ${\rm Im}\,u_{--}$ in Eq.~(\ref{vacua}) at the order
considered here. As $z\to \infty$ the solution $u_2(z)$ grows exponentially
unless the relation
\begin{equation}
\int_{-\infty}^\infty\!\!{\rm d}x
\,\frac{-\nu_2 +\alpha\left[\nu_{1+}^2-(u^{(0)}(x))^2\right]
-\mu_2\left[u^{(0)}(x)-v^{(0)}(x)\right]}
{\cosh^2\left[\nu_{1+}\,m\,(x-z_u)\right]}=0
\label{sepfix}
\end{equation}
is fulfilled. Once this relation is met the $z\to \infty$ boundary
condition $u_2\to {\rm Im} \, u_{++}$ is also satisfied.

The relation (\ref{sepfix}) can be viewed as a constraint
ensuring orthogonality of the
inhomogeneous part in the $u_2$-equation and the zero mode
$\cosh^{-2}\left[\nu_{1+}\,m\,(z-z_u)\right]$
in $u_1$. This approximate zero mode corresponds to a shift of the
$u$-soliton center and at the same time also represents the spatial
dependence of the corresponding fermionic zero
mode~\footnote{In the next section we will show that this orthogonality
condition is equivalent to the vanishing of a particular
supercharge.}. 
The relation
(\ref{sepfix}) then fixes the separation $z_u-z_v$ 
of the two primary solitons and can be presented in the form:
\begin{equation}
w_{\nu_1}(z_u-z_v)=\frac{\nu_2}{\mu_2\, \ka}\;,
\label{condition}
\end{equation}
where
\beq
\kappa=\frac{1}{2}\left[\nu_{1+}^{3}+\nu_{1-}^{3}\right]~,
\label{kappa}
\eeq
and the function $w_\nu$ of the soliton separation $s=z_u - z_v$ is
defined as
\begin{equation}
w_\nu(s)=\frac{1}{2}\int_{-\infty}^\infty \frac{{\rm d}x }{\cosh^2 x}
\,
\tanh\left[\frac{\nu_-}{\nu_+}\,x+\nu_-\,m\,s\right]~.
\label{wnuofs}
\end{equation}

It is important that the condition (\ref{condition}) also ensures that
the solution for $v_2$,
\begin{eqnarray}
v_2(z)&=&\cosh^2\left[\,\nu_{1-}\,m\,(z-z_v)\right]\times
\nonumber\\[0.1cm]
&&m\int_{-\infty}^z\!\!{\rm d}x
\,\frac{\nu_2 +\alpha\left[\nu_{1-}^2-(v^{(0)}(x))^2\right]
+\mu_2\left[u^{(0)}(x)-v^{(0)}(x)\right]}
{\cosh^2\left[\nu_{1-}\,m\,(x-z_v)\right]}~,
\end{eqnarray}
is finite at both $z \to +\infty$ and $z \to -\infty$, and thus
satisfies the proper boundary conditions. As for the solutions for the
real parts, $u_1$ and $v_1$, described by the first pair of equations in
(\ref{bps-mu}), these solutions always exist, due to the existence of
real BPS solitons in the model with real 
parameters \cite{Shifman:1998wg},
as discussed in Sec.~\ref{sec:3.1}. Thus no additional constraint arises.

\begin{figure}[ht]
     \epsfxsize=8cm
\centerline{     \epsfbox{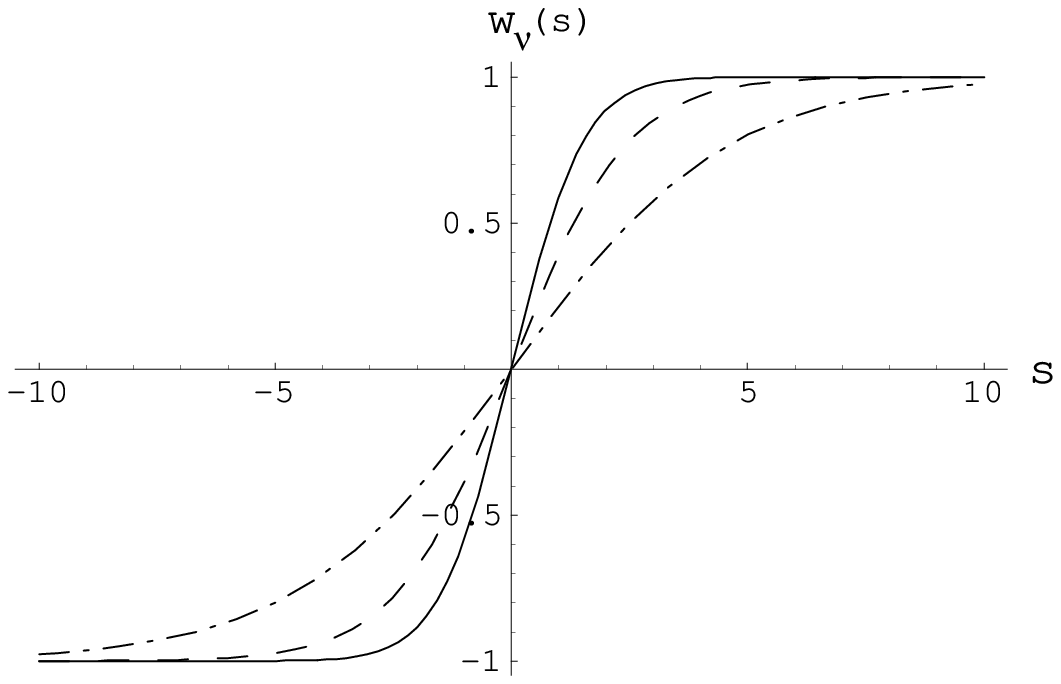}}
    \newcaption{Plots of $w_\nu(s)$ at $\nu=0$ (solid), $\nu=0.8$
(dashed),
$\nu=0.95$ (dot-dashed), $s$ is measured in units of $1/m$.
\label{fig:cmsw1}}
\end{figure}

Here we make a few remarks on the properties of the function $w_\nu(s)$,
defined by the integral (\ref{wnuofs}). The symmetry properties of this
function can easily be seen by writing it as the derivative 
$w_\nu(s) ={\rm d} g(s)/{\rm d} s$ of the function
\beq
g_\nu(s)={m \over 2} \, \int_{-\infty}^\infty {\rm d} z \left \{ 1-
\tanh \left [\nu_+ \,m\, z \right] \, \tanh \left [\nu_- \,
m\,(z+s) \right] \right \}~,
\label{gofs}
\eeq
which is symmetric under separate reversal of the sign of $\nu$ or $s$.
Thus $w_\nu(s)$ is even in the
index $\nu$: $w_{-\nu}(s)=w_\nu(s)$, and is odd in the variable $s$:
$w_\nu(-s)=-w_\nu(s)$, and is monotonically increasing from
$w_\nu(s\to-\infty)=-1$ to $w_\nu(s\to+\infty)=+1$. At
large positive $s$ its asymptotic behavior is given by
\beq
w_\nu(s)=1-{ \nu_++\nu_- \over \nu} \left[ \,\nu_+\,
\exp(-2 \, \nu_- \,m\, s) - \nu_- \, \exp(-2 \, \nu_+ \,
m\,s) \right]+ \cdots ~,
\label{wass}
\eeq
where the ellipses stands for higher powers and mixed products of the
two exponents: $\exp ({-2 \, \nu_- \, m\,s})$ and 
$\exp ({-2 \,\nu_+ \,m\, s})$.
At $\nu=0$ the integral in equation (\ref{wnuofs}) can be expressed in
terms of elementary functions,
\begin{equation}
w_0(s)= \coth m\,s -\frac{m\,s}{\sinh^2 m\,s}\;,
\label{w0ofs}
\end{equation}
and the asymptotic behavior of $w_0(s)$ as $s  \to +\infty$:
\beq
w_0(s)=1-(4 \,m\, s +2) \, e^{-2 \, m\,s} + \ldots
\label{w0ass}
\eeq
is in agreement with the $\nu \to 0$ limit of the expression
(\ref{wass}). Plots of $w_\nu(s)$ for a few values of $\nu$ are shown in
Fig.~\ref{fig:cmsw1}.

The limited magnitude of $w_\nu(s)$, $|w_\nu(s)|\leq 1$, means that the
BPS solution we consider only exists in the range
\begin{equation}
\left|\nu_2\right|\leq\left|\mu_2\right|\,\kappa \;.
\end{equation}
As expected the boundaries of this range coincide with the part of CMS found
previously by algebraic means near the real $\nu$ axis.

It is then simple to find the value for the distance $s_0$ between the
primary solitons in the BPS composite state. Say, for
$\nu_1=0$, we have when $|\mu_2|-|\nu_2|\ll |\mu_2|$
\begin{equation}
 {\rm e}^{m\,|s_0|} = \et\, \ln \eta, \quad{\rm where}\quad
 \eta = \sqrt{\frac{|\mu_2|}{|\mu_2|-|\nu_2|}}.
\end{equation}

\section{Quantum Mechanics of Two Solitons}
\label{sec:four}

The BPS state which connects the vacua $\{++\}$ and $\{--\}$ and exists
within the stability domain can be viewed as a bound state of one
$u$-soliton, located at $z=z_u$, and one $v$-soliton, located at $z=z_v$. The
equilibrium separation between the solitons, $s=z_u-z_v$, at which the
minimum is achieved at given $\nu_2$ and $\mu_2$, is determined from the
equation (\ref{condition}). In this section we consider the supersymmetric
quantum mechanics of the two soliton system. The SQM system 
describes the BPS bound state (which is the ground state in the
problem) within the stability domain, as well as low-lying non-BPS
exited states.

The formulation of this problem refers to an effective description of
the two solitons as heavy particles with masses $M_u$ and $M_v$ in
terms of their coordinates $z_u$ and $z_v$. This approximation
is natural near the CMS where the binding energy is small relative to
the soliton masses. For slowly moving solitons
$|\dot z_u|, |\dot z_v| \ll 1$ the nonrelativistic energy can be written
as
\beq
E  = M_u \, {\dot z_u^2 \over 2}+ M_v \, {\dot z_v^2
\over 2} + U(s)~,
\label{energy2}
\eeq
where the dot denotes the time derivative, and $U(s)$ is the interaction
potential depending on the separation $s=z_u-z_v$ between the solitons.
Separating out the center of mass coordinate, we come to a standard
quantum mechanical Hamiltonian for the relative motion,
\begin{equation}
H=\frac{p^2}{2M_r}+U(s)\;.
\end{equation}
The supersymmetric generalization of this Hamiltonian is given in
Eq.~(\ref{hsqm}) which depends on the superpotential $W(s)$.
Below we will find the expression for this SQM
superpotential by comparing the field theoretic supercharges
evaluated on the soliton solutions
with the SQM realization in Eq.~(\ref{reali}).
An alternative derivation of $W'$ based solely on
conventional bosonic considerations is presented in an Appendix.

\subsection{SQM superpotential from field-theoretic supercharges}
\label{sec:4.1}

The action (\ref{uvact}) for the Wess-Zumino model
leads to the following expression for the supercharges $Q_\alpha$,
\begin{equation}
Q=\frac{m^2}{\sqrt{2}\,\lambda^2}\int \!{\rm d} z\left[ \partial_t
u\;\psi
+\partial_z u\;
\gamma^0\gamma^1\psi +i\,m\,\partial_{\bar u} \overline{ {\cal W}}
\;\gamma^0
\psi^* +(u\to v\,, ~\psi \to \eta)
\right]\;,
\label{scha}
\end{equation}
 where $u$, $\psi_\alpha$ and  $v$, $\eta_\alpha$ are the bosonic and
fermionic components of the $U$ and $V$ superfields, and ${\cal W}(u,v)$ is the
superpotential of the model. The two remaining supercharges  
$\bar Q_\alpha$ are just the complex conjugates of $Q_\alpha$.

Let us first evaluate the supercharges for the $u$-soliton in the
leading approximation, i.e. when $\nu_2\!=\!\mu_1=\!\mu_2\!=\! 0$. The field
$u$ is given by Eq.~(\ref{lead}),
$u=u^{(0)}=\nu_{1+}\tanh\left[\nu_{1+}\,m\,(z-z_u)\right]$,
while $v$ is a constant, $v=-\nu_{1-}$. For the fermionic fields we
substitute zero modes, two of which are in the field $\psi_\alpha$,
and there are none in $\eta_\alpha$,
\begin{equation}
\psi_{\rm zero~modes}=\left(\!\!\begin{array}{c} i\,b_u\\ a_u
\end{array}\!\!\right)\frac{1}{\sqrt{2M_u}}\,\partial_z u^{(0)}.
\label{ferm}
\end{equation}
In this expression 
$M_u=(4m^3/3\lambda^2)\nu_{1+}^3$ is the mass of the $u$-soliton,
$a_u$ and $b_u$ are real fermionic operators entering as coefficients of
the normalized zero modes, and their algebra is fixed 
by canonical quantization,
\begin{equation}
a_u^2=b_u^2=1\;,\quad \{a_u,b_u\}=0\;.
\label{aubu}
\end{equation}
Upon these substitutions, the supercharge $Q_\alpha$ becomes
\begin{equation}
Q=\sqrt{M_u}\left(\!\!\begin{array}{c} a_u\\i\,b_u
\end{array}\!\!\right)\;,
\end{equation}
which can be rewritten in terms of real charges (see
Eq.~(\ref{realch}) in Sec.~\ref{sec:2.3} for definitions),
\begin{equation}
Q^1_1=\sqrt{2M_u}\; a_u\;, \quad Q^2_2=\sqrt{2M_u}\; b_u\;, \quad
Q^1_2=Q^2_1=0\;.
\label{qu}
\end{equation}
The  result for the supercharges matches the general construction of
Sec.~\ref{sec:2.3} wherein the operators $a_u$ and $ b_u$ can be 
realized as Pauli 
matrices, e.g. $a_u=\tau_1$ and $b_u=\tau_2$.

Now let us find the supercharges corresponding to the $\{1,1\}$
configuration of the $u$- and $v$-solitons at
$\nu_2\!=\!\mu_1=\!\mu_2\!=\! 0$. We choose boosted soliton solutions,
\begin{eqnarray}
&&
u^{(0)}=\nu_{1+}\tanh\left[\nu_{1+}\,m\left(z-z_u-\frac{p}{M_u}\,t\right
)\right]\;,
\nonumber\\[2mm]
&&
v^{(0)}=\nu_{1-}\tanh\left[\nu_{1-}\,m\left(z-z_v+\frac{p}{M_v}\,t\right
)\right]
\;,
\end{eqnarray}
where $p$ is their relative momentum, and the total momentum is zero.
The fermions are given by Eq.~(\ref{ferm}) for $\psi$ and by a
similar expression for $\eta$ with the substitution $u\to v$,
where $u$- and $v$-fermions anticommute. With 
time-dependent solutions the terms 
$\partial_t u \,\psi$,  $\partial_t v\, \eta$ now contribute to 
the supercharges~(\ref{scha}),
\begin{eqnarray}
&& \!\!\!\!\! \!\!Q^1_1=\sqrt{2(M_u+M_v)}\; (a_u\cos \delta +a_v\sin
\delta)\;,
\quad Q^2_2=\sqrt{2(M_u+M_v)}\; (b_u\cos \delta +b_v\sin \delta)\;,
\nonumber\\[2mm]
&& \!\!\!\!\! \!\! Q^1_2 =\frac{p}{\sqrt{2M_r}}\; (-a_u\sin\delta
+a_v\cos \delta)
\;, \qquad \quad ~~Q^2_1=\frac{p}{\sqrt{2M_r}}\; (-b_u\sin\delta
+b_v\cos
\delta)\;,
\label{quv}
\end{eqnarray}
where we have defined 
$\cos \delta=\sqrt{M_u/(M_u+M_v)}$. We observe that the relative motion
implies that the `composite' state is non-BPS in the absence of 
any interaction between the solitons.

In order to switch on the interaction we consider nonzero $\mu$ and
$\nu_2$.  To obtain the result to first order in these parameters 
it is enough to substitute the same leading order expressions
for the bosonic and fermionic fields accounting for the terms linear
in  $\mu$ and $\nu_2$ in the expression (\ref{scha}) for the supercharges, 
as well as for the phase $\alpha$ of the central charge. The linear
dependence on $\mu$ and $\nu_2$ arises from the terms
\begin{equation}
\frac{m^2}{\sqrt{2}\,\lambda^2}\int \!{\rm d}z
\left[\mu\;(u^{(0)}-v^{(0)})-i\nu_2\right] \gamma^0 (\psi^* -\eta^*)
\end{equation}
in (\ref{scha}). The phase  $\alpha$
is also linear in $\nu_2$ (see Eq.~(\ref{alpha})), and needs to be
taken into account in Eq.~(\ref{realch}) when relating $Q_\alpha$ with
$Q^{1,2}_\alpha$.

The resulting supercharges are ($ Q^1_1$,  $Q^2_2$ are not changed and
are written here for completeness)
\begin{eqnarray}
&& Q^1_1=\sqrt{2(M_u+M_v)}\; a\;,
\qquad \qquad ~~Q^2_2=\sqrt{2(M_u+M_v)}\; b\;,
\nonumber\\[2mm]
&& Q^1_2 =\frac{1}{\sqrt{2M_r}}\left[p\; \tilde a + W'(s)\; \tilde b\right]
\;, \qquad \quad ~~Q^2_1=\frac{1}{\sqrt{2M_r}}\left[p\; \tilde b - 
W'(s)\; \tilde a\right]
\;,
\label{quvmu}
\end{eqnarray}
where we denote
\begin{eqnarray}
&& a=a_u\cos \delta +a_v\sin\delta\;,\quad ~~b=b_u\cos \delta
+b_v\sin\delta\;,
\nonumber\\[1mm]
&&\tilde a=-a_u\sin\delta +a_v\cos \delta\;,\quad \tilde
b=-b_u\sin\delta
+b_v\cos \delta\;.
\end{eqnarray}

The quantum-mechanical superpotential (more precisely its
derivative $W'$) is then read from $Q^1_2$ (or $Q^2_1$) to be
\begin{equation}
W'(s)= \frac{3M_r}{1-\nu_{1}^2}\left[ \,\mu_2\,\kappa\,
w_{\nu_1}(s)-\nu_2\,\right]\;,
\label{suppote}
\end{equation}
where
\begin{equation}
M_r=\frac{M_u
M_v}{M_u+M_v}=\frac{2}{3}\,\frac{m^3}{\lambda^2}\,\frac{(1-\nu_{1}^2)^{3
/2}}
{\kappa~}\;,\qquad \kappa=\frac{1}{2}\,(\nu_{1+}^3+\nu_{1-}^3)\;,
\label{reduce1}
\end{equation}
and the function $w_{\nu}(s)$ is defined by Eq.~(\ref{wnuofs}).

The SQM Hamiltonian then has the form
\begin{equation}
H_{\rm SQM}=(Q^1_2)^2=(Q^2_1)^2=\frac{1}{2M_r}\left[p^2 + (W'(s))^2
-i\,W''(s)\,\tilde a\,\tilde b\right]\,.
\end{equation}

An explicit matrix realization of the four operators $a_{u,v}$ and
$b_{u,v}$, satisfying the Clifford algebra, can a priori be 
chosen in the factorized form: $a_u\!=\!\tau_1 \otimes\sigma_3$, 
$b_u\!=\!\tau_2 \otimes \sigma_3$, $a_v\!=\!I \otimes \sigma_1$, 
and $b_v\!=\!I \otimes \sigma_2$. This factorized form of
the fermionic operators realizes a description in terms of two independent
particles. This choice is perfectly acceptable and realizes the 
\ntwo superalgebra (\ref{alg3}).  However, it differs from 
the specific realization (\ref{reali}) by an orthogonal rotation of 
angle $\delta$. In order to match the conventions used in Eq.~(\ref{reali})
for a description of the two-soliton system, one has to use an equivalent
representation of these operators, obtained by the inverse rotation:
\begin{eqnarray}
&&a_u = \tau_1 \otimes \sigma_3 \, \cos \delta - I \otimes \sigma_1 \,
\sin \delta \,,~~~~b_u=\tau_2 \otimes \sigma_3 \, \cos \delta - I
\otimes \sigma_2 \, \sin \delta \,, \nonumber \\
&&a_v = \tau_1 \otimes \sigma_3 \, \sin \delta + I \otimes \sigma_1 \,
\cos \delta \,,~~~~b_v=\tau_2 \otimes \sigma_3 \, \sin \delta + I
\otimes \sigma_2 \, \cos \delta \,.
\label{mtr}
\end{eqnarray}

The final expression for the full quantum Hamiltonian of the two-soliton
system can be written as
\beq
H_{\rm SQM} = {p^2 \over 2 \, M_r}+ {9\,M_r \over 2} \, \frac{\left[\,
\mu_2 \, \kappa \,
w_{\nu_1}( s)-\nu_2\,\right]^2}{  (1- \nu_{1}^2)^2}+{3 \over 2} \,
{\mu_2 \, \kappa \, w'_{\nu_1} ( s) \over (1-\nu_1^2)}\,\sigma_3
\label{hintq}
\eeq
with $M_r$ given in Eq.~(\ref{reduce1}) and we use the matrix
representation~(\ref{mtr}) for the fermions (omitting the tensor
product with unity in $H_{\rm SQM}$ ).

It is worth noting a couple of limits in which the potential 
simplifies, and can be expressed in terms of elementary functions.
Recall first of all that when $\nu_1=0$ the function
$w_{0}(s)$ is known analytically (see Eq.~(\ref{w0ofs})). If $\nu_1$ is
not too large, i.e. $\nu_1 \leq 0.5$, there also exists a convenient 
simplified form in which the superpotential is very closely 
approximated by the expression
\be
 W'_{\rm approx}(s)  = 3M_r\,\left[\,\mu_2
      \tanh(ms)-\nu_2\,\right],
\ee
where the reduced mass $M_r$ is taken at $\nu_1=0$.  In this
superpotential we recognize the simplified model discussed in 
Sec.~\ref{sec:2.2}
(see Eq.~(\ref{toysp})).
Another simple case arises when $\nu_1^2$ is close to 1, i.e.
$1-\nu_1^2\ll 1$,
\begin{equation}
W'= \frac{3M_r}{
\sqrt{1-\nu_1^2}}\left(\mu_2\,
   m\,s-\nu_2\right).
\end{equation}
The potential in this case reduces to that of the harmonic oscillator.

In the limit of large separation, the potential energy in the
Hamiltonian~(\ref{hintq}) tends to a constant which depends in the
sign of $s$,
\begin{equation}
U_\pm=U(s\to \pm\,\infty)={9\,M_r \over 2} \, \frac{\left(
\pm\,\mu_2 \, \kappa -\nu_2\right)^2}{  (1- \nu_{1}^2)^2}\;,
\label{upm}
\end{equation}
(Note, however, that the spin dependent $\sigma_3$ term does not contribute).
These constants denote the energy levels at which the continuum states appear
while the ground state, which is the
$\{1,1\}$ BPS soliton, is a zero energy eigenfunction of $H_{\rm SQM}$.

The origin of the two continuum thresholds is that at nonzero $\mu$
the classification for solitons we introduced at $\mu=0$ is no longer
sufficient ---
the $u$-soliton interpolating between the $\{-+\}$ and $\{++\}$ vacua (see
Fig.~\ref{fig:vac}) is different from the $\tilde u$-soliton interpolating
between the $\{--\}$ and $\{+-\}$ vacua, and a similar
distinction arises between the $v$- and $\tilde
v$-solitons. Thus, the system under consideration at large $s$ describes two
channels: the $u$ plus $v$ solitons at positive $s$, and the 
$\tilde u$ plus $\tilde v$ at negative $s$.
It is straightforward to verify this by calculating the two binding energies,
\begin{eqnarray}
&&\!\!\!\!\!\!\!\!\Delta E_+=M_{1,1}-M_u - M_v= {m^3 \over \lambda^2} \,
\left [
\,|{\cal W}_{++}-{\cal W}_{--}| -|{\cal W}_{++}-{\cal W}_{-+}| -|{\cal
W}_{-+}-{\cal W}_{--}|\,
\right],\nonumber\\[-1mm]
&&\\[-2mm]
\label{deltae}
&&\!\!\!\!\!\!\!\!\Delta E_-=M_{1,1}-M_{\tilde u} + M_{\tilde v} =
{m^3\over\lambda^2} \,
\left [
\,|{\cal W}_{++}-{\cal W}_{--}| -|{\cal W}_{+-}-{\cal W}_{--}| -|{\cal
W}_{++}-{\cal W}_{+-}|
\,
\right],\nonumber
\end{eqnarray}
from which we observe that $\Delta E_\pm=-U_\pm$.
Note that, although the quantities $\Delta E_\pm$ are of
second order in $\mu_2$ and $\nu_2$, it is sufficient to use
the expressions (\ref{vacua}) which are only valid to first order in
$\mu$ (and are formally exact in $\nu$) for the values  of
${\cal W}_{ij}$. This is due to the fact that for
real $\nu$ and $\mu$ the values of ${\cal W}_{ij}$ are real and $\Delta
E_\pm$ vanishes. Thus $\Delta E_\pm$ arises as an effect quadratic in
the imaginary parts of the differences of ${\cal W}_{ij}$ which by
themselves are linear in $\nu_2$ and $\mu_2$.

Finally, we also write down the asymptotic behavior
of the potential as $s\to \pm\, \infty$,
\be
 U(s) \; \longrightarrow \; U_{\pm} + K_{\pm} \exp(-2\nu_{1-}\, m |s|) 
+ \cdots,
\ee
where the coefficient of the leading exponential term is
\be
 K_{\pm} \; = \;  \frac{3\,(\nu_{1+}+\nu_{1-})\,\mu_2 
\kappa}{\nu_1\nu_{1+}(1-\nu_1)(1-\nu_1^2)}\!
\left[ -6M_r\,(\mu_2 \kappa \mp
\nu_2)+m\,\sigma_3\,(1-\nu_1^2)\,\nu_{1-}\right].
\ee
We have made the assumption here that
$\nu_{1-}<\nu_{1+}$. We see that the characteristic distance 
$s$ is defined by $1/m\nu_{1-}$ which as expected is the wavelength for the
lightest particle in the model. We also observe that 
the spin dependent term contributes
to the exponential tail. Moreover, on the CMS where $\mu_2 \kappa \mp\nu_2=0$,
it is the only contribution. This `fermionic dominance' takes place 
in a very narrow region near the CMS,
\begin{equation}
\left|\mu_2 \kappa \mp\nu_2\right|\ll
\frac{(1-\nu_1^2)\,\nu_{1-}}{6}\,\frac{m}{M_r}\;.
\label{domin}
\end{equation}
The effect of this regime of enhanced quantum corrections near the
CMS will be considered in more detail in the next subsection.

\subsection{Properties of the two-soliton system}

As expected, the second term in the potential in Eq.~(\ref{hintq})
is of higher order than the first in the loop expansion parameter
$\lambda$. However the second term is of lower order in the small
parameter $\mu_2$
and, by tuning $\mu_2$, one can study this potential both in the
classical limit corresponding to $\mu_2 \gg \lambda^2/m^2$ and in the
quantum limit $\mu_2 \ll \lambda^2/m^2$, or anywhere in between as long
as the condition for validity of the formula (\ref{hintq}), $\mu_2
\ll 1$, is maintained. Upon a slightly more detailed inspection of
classical {\em vs} quantum effects in the Hamiltonian
(\ref{hintq}) one can readily see that in fact the quasiclassical
parameter in this system is not just the ratio $\lambda^2/(m^2 \mu_2)$
but is determined by the parameter
\be
 \xi = \frac{W''''(s_0)}{(W''(s_0))^2},
\ee
introduced in Sec.~\ref{sec:two}, 
where $s_0$ is the classical equilibrium separation 
determined by (\ref{condition}). Recall that $\xi$ measures the
quantum correction to the curvature of the potential near the classical
minimum, the system being essentially classical for $\xi\ll 1$, and
highly quantum for $\xi\gg 1$.

For the model at hand we find
\beq
\xi = \frac{w'''_{\nu_1}(s_0)}{2\,m\, \mu_2\,\sqrt{1-\nu_1^2}
\left(w'_{\nu_1}(s_0)\right)^2}\cdot {\lambda^2 \over   m^2 }\;.
\label{xiq}
\eeq
Near the CMS the equilibrium distance $s_0$ is large, and 
$\xi$ takes the form
\beq
\xi = \frac{ \kappa}{\nu_{1+}\,( \,\mu_2\,
  \kappa\mp \nu_2)} 
\cdot  {\lambda^2 \over   m^2 }\;,
\label{xiqas}
\eeq
where for definiteness we have again assumed that $\nu_{1-}<\nu_{1+}$. Notice
that the condition $|\xi|\gg 1$ agrees with 
Eq.~(\ref{domin}) which, as discussed above, 
defines the essentially quantum regime in the
narrow region along the CMS.

\begin{figure}[ht]
 \epsfxsize=14cm
\centerline{\epsfbox{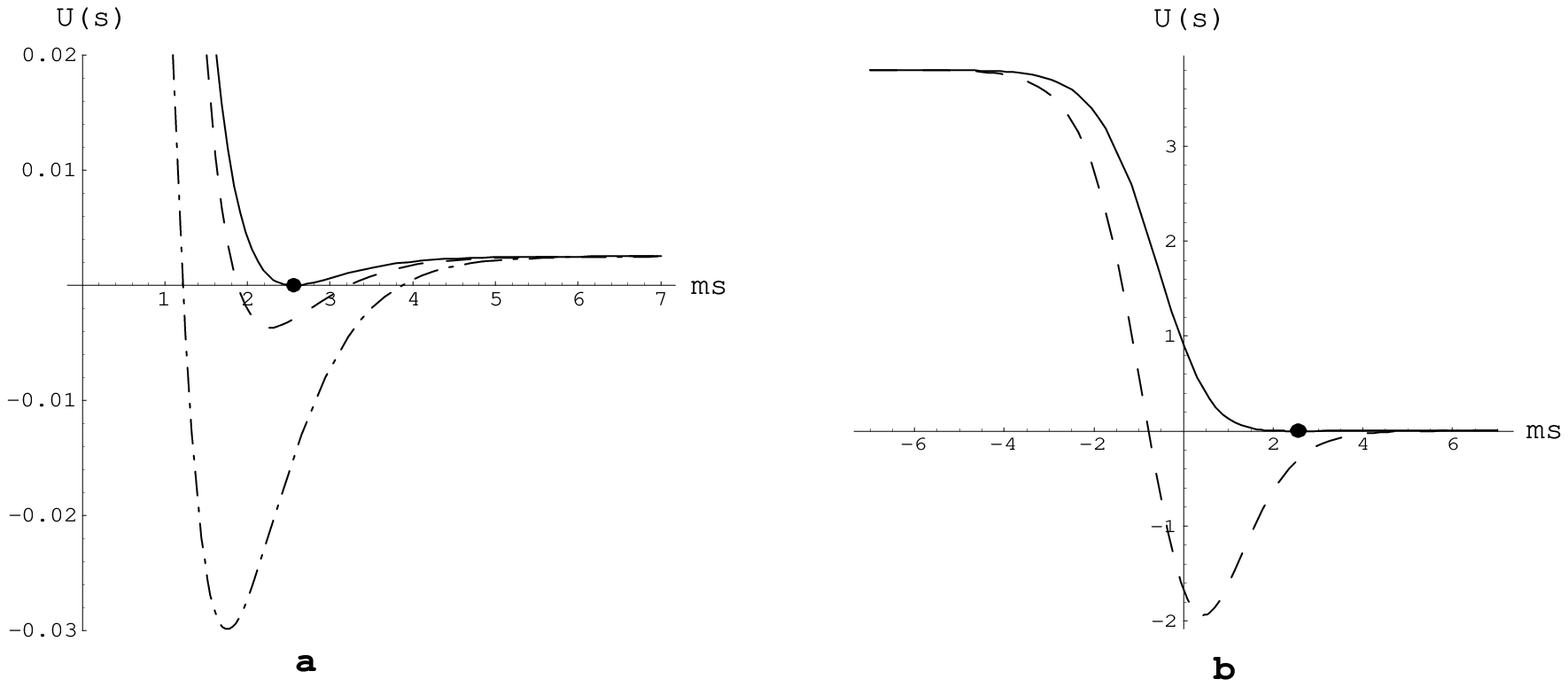}}
\newcaption{Plots of the full potential $U(s)$ (arbitrary units) 
at $\nu_1=0$, $\nu_2/\mu_2\!\!=\!\! 0.95$ for several values of $\xi$.
The classical equilibrium point is at
$m s\, \approx\! 2.56$ and is shown by heavy dot. Fig.~a: details of
the potential near minimum for $\xi=0.009$ (solid), $\xi=0.9$ (dashed), and
for $\xi=4.45$ (dot-dashed). Fig.~b: the potential shown at a larger scale. The
curves for  $\xi \le 4.45$ are practically unresolvable and coincide with
the solid curve, the dashed curve corresponds to $\xi=125$. It can
be noticed that the latter value of $\xi$ still corresponds to moderate
values of $\lambda/m$: $\lambda^2/m^2 \approx 7.7 \,
\mu_2$.\label{fig:pot}}
\end{figure}

\begin{figure}[ht]
\epsfxsize=7cm
\centerline{\epsfbox{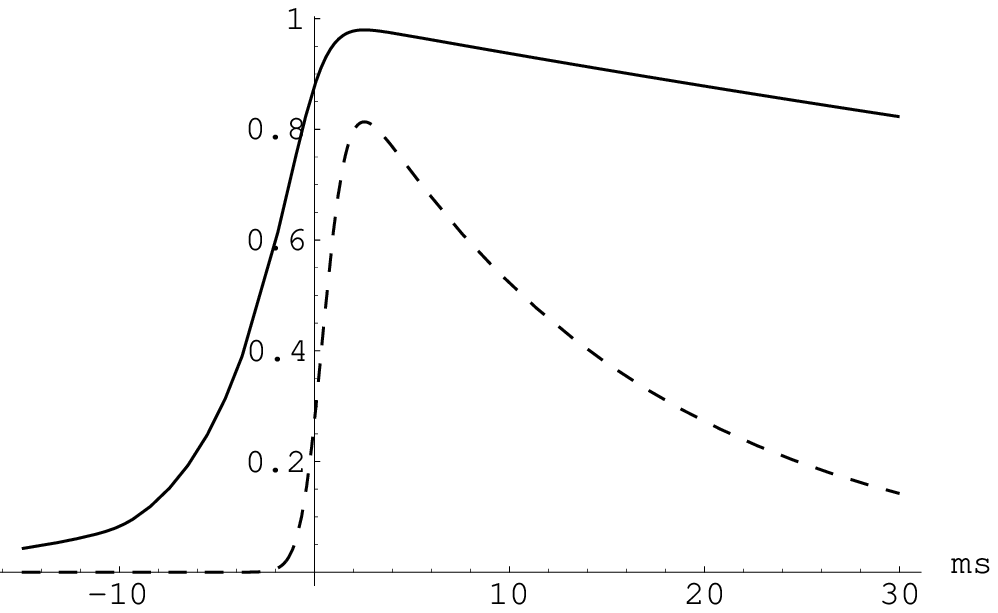}}
\newcaption{Plots of the ground state wave function $\ps_0(s)$ for
$\xi=17.7$ (dashed) and $\xi=177$ (solid). As above, these
parameters correspond to $\nu_2/\mu_2=0.95$ and the classical
equilibrium point is at $ms\approx 2.56$.\label{fig:wfn}}
\end{figure}

When the system admits a supersymmetric ground state,
the corresponding wave function $\psi_0(s)$ can always be found as
\beq
\psi_0(s)={\rm const} \, \exp \left [ -  W(s) \right]
=
{\rm const} \, \exp \left [ -3\,\frac{M_r}{m}\;
\frac{\,\mu_2\,\kappa\,
g_{\nu_1}(s)-\nu_2\,ms}{1-\nu_{1}^2}\right]~,
\label{psiofs}
\eeq
where for definiteness we again assume $\mu_2 > 0$, and $g_\nu(s)$,
defined by Eq.~(\ref{gofs}), is the integral of $w_\nu(s)$.
Independently of the quasiclassical parameter $\xi$ the maximum of
$\psi_0(s)$ is always located at $s_0$. However the spread of the wave
function, i.e. the dispersion of the distance between the solitons in
the BPS bound state, essentially depends on the parameter $\xi$. As $\xi
\to 0$ the full potential has a minimum at $s=s_0$, and the system is
classically located at the minimum. At larger $\xi$ the minimum of the
full potential shifts towards $s=0$, reaching $s=0$ in the limit $\xi
\gg 1$, but the maximum of the wave function is still at $s=s_0$. In
the latter extreme quantum limit the system resembles the deuteron: the
wave function spreads over distances much larger than the size of the
interaction region. In the two-soliton system this behavior is even
more drastic at large $\xi$ than in the deuteron: the wave function
reaches its {\it maximum} far beyond the interaction region. The
classical and the quantum behavior of the system at different values of
$\xi$ is illustrated by a series of plots in
Fig.~\ref{fig:pot},~\ref{fig:wfn}.
One may also note that in general the interaction of the two
solitons is strong only at distances of order $m^{-1}$ near $s=0$: the
potential is asymmetric in $s$ and changes rapidly near $s=0$, i.e. the
force is strongest when the solitons substantially overlap. 
In a narrow region near the CMS, given by Eq.~(\ref{domin}), an attraction 
at short distances  creates an essentially
quantum state, resembling a deuteron.
Deeper into the stability region an exponentially shallow minimum of the
potential at large $s_0$ results in a quasiclassical bound state.

Once one crosses the CMS, the wave function (\ref{psiofs}) is no
longer normalizable, and the physical ground state of the system
is non-supersymmetric. The broadening of the wave function for the
bound state near the CMS is exhibited in Fig.~\ref{fig:wfn}.
Thus, as discussed in Sec.~\ref{sec:2.2} the bound state
level reaches the continuum on the CMS, where it completely delocalizes,
and on crossing the CMS the $\{1,1\}$ bound state is no longer
present in the physical spectrum.

\subsection{Another dynamical regime: Extra Moduli on the CMS}
\label{sec:4.3}

The scenario discussed above, involving a short range superpotential
which remains finite on the CMS, is only a generic description for
the near CMS dynamics in certain systems. As discussed
in Section~2, a different dynamical scenario is possible
if there exist extra moduli on the CMS.
However, it turns out that the two-field model also exhibits a
dynamical regime of this type, and we observe in this case
that the approach to the CMS is still characterized 
by delocalization of the bound state wavefunction, albeit in a 
somewhat different manner to the case
considered above. 

\begin{figure}[ht]
\epsfxsize=8cm
 \centerline{%
   \epsfbox{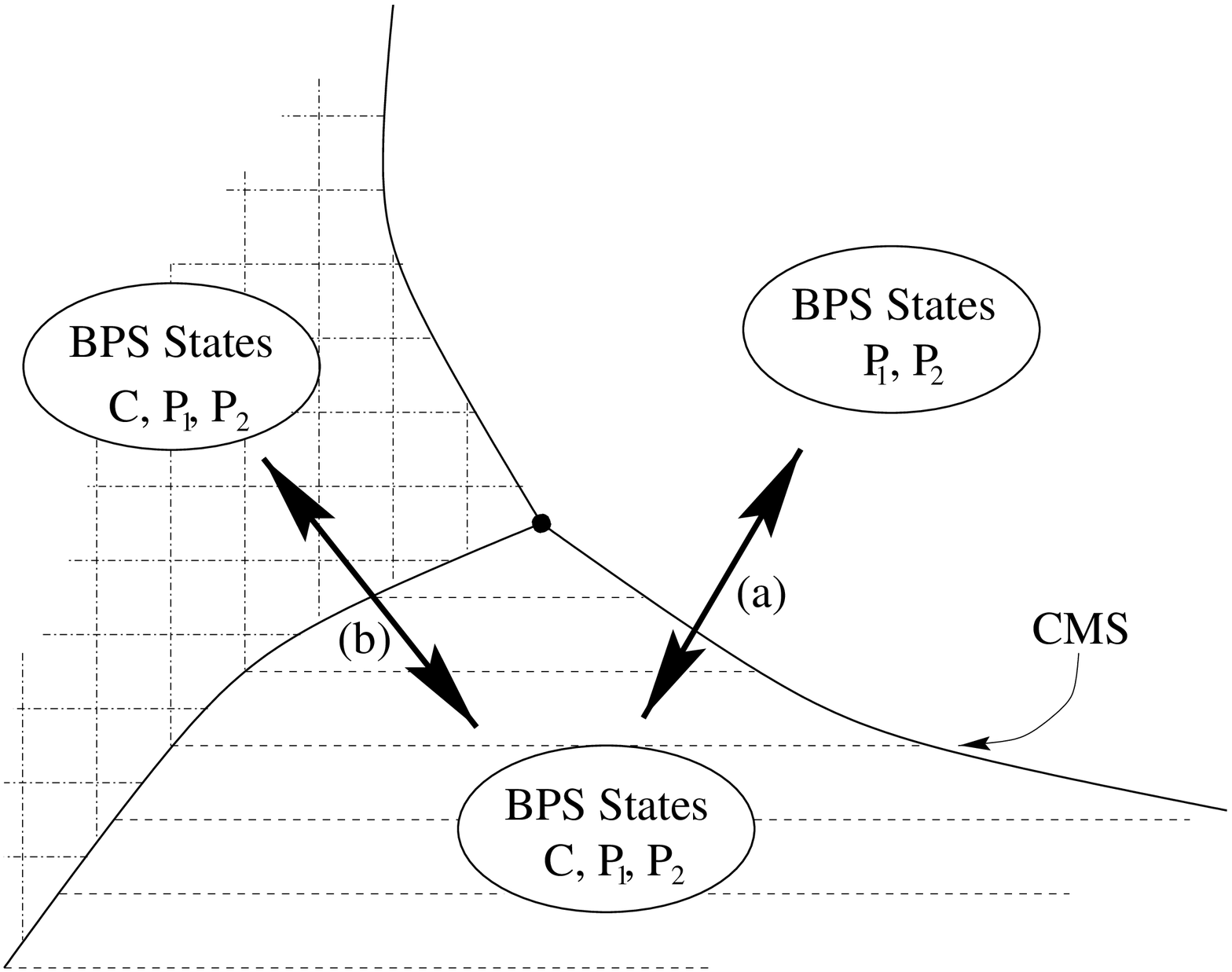}%
         }
 \newcaption{Possible scenarios for the BPS spectrum, taken from a
small region of Fig.~\ref{fig:cmsf2}, where $P_1$ and
$P_2$ refer to the two primary solitons, while $C$ refers to
the composite kink: (a) A composite bound state $C$ exists only on one
side of the CMS; (b) a bound state exits on both
sides of the CMS.\label{fig:cms_ie}}
\end{figure}

Firstly, recall that
in the example considered above with $\mu_2\neq 0$, the approach
to the CMS was determined by Eq.~(\ref{condition}). In the `interior'
domain, $|\nu_2|<|\mu_2|\ka$, the equation $W'(s)=0$ has a solution and
consequently the composite soliton was BPS saturated. Upon approach
to the CMS, the zero of $W'(s)$ runs to infinity and, after
crossing the CMS at $|\nu_2|>|\mu_2|\ka$, there is no longer a solution
to $W'(s)=0$ and hence no BPS soliton. This scenario is illustrated by
see Fig.~\ref{fig:cms_ie}a..

Now we consider a different dynamical regime, see Fig.~\ref{fig:cms_ie}b,
where, in both the `interior' and `exterior' regions, the
spectrum of BPS states is the same (although possibly rearranged). In
this case one still has delocalization on the CMS one still has
delocalization, although only
for the wave function in this case as there is no diverging
(classical) separation of the constituents. 
To this end let
us set $\nu=0$ (i.e. discard the term linear in $U,V$ in the
superpotential (\ref{wphix})). As explained in Sec.~\ref{sec:3.1}, at $\nu=0$
the CMS is very simple,
\be
 {\rm Im}\,\mu\equiv \mu_2 = 0.
\ee
The SQM system (\ref{wqm2}) one arrives at in this case is described by
the superpotential,
\be
 W'(s) = 2 \,\frac{m^{3}}{\la^2}\,\mu_2 \,w_0(s),
\ee
where $w_0(s)$ was defined in (\ref{w0ofs}). For our purposes it is
important that $w_0(0)=0$, and that there are no other zeros of $w_0(s)$. 
Thus the solitons always overlap classically. However, as one approaches the
CMS the wave function still spreads out due to the fact that
$\mu_2\rightarrow 0$. In particular, at large $|s|$,
\be
 w_0(s) \longrightarrow ~{\rm sgn}(s),
\ee
and one observes that the zero energy bound state exists for both
positive and negative $\mu_2$ (see Fig.~\ref{fig:cms_ie}b). The wave function
at large $s$ is
\be
 \exp\left(2\frac{m^3}{\la^2}\mu_2 s\si_3\right)
\ee
times either $|\downarrow\rangle$ or $|\uparrow\rangle$ depending on the
eigenvalue of $\si_3$. As $|\mu_2|\rightarrow 0$ the bound state level
approaches the continuum spectrum while the wave function swells. At
$\mu_2=0$ the wave function is completely delocalized and there
is no binding.

As alluded to above, this
dynamical regime is distinct from that considered previously
where $W'$ remained finite on the CMS; rather the CMS was characterized
by the escape of the root of $W'(s)=0$ to infinity. In contrast, in the
example considered here the root of the equation $W'(s)=0$
does not shift at all. Despite this one may note that $\xi$ still
diverges near the CMS due to its inverse dependence on $\mu_2$.

In fact, precisely on the CMS the potential vanishes, and thus a 
new quantum modulus arises corresponding to the relative separation of the
constituents.

\section{Dyons in SU(3) ${\cal N}=2$  SYM}
\label{sec:five}

We turn now to consider similar phenomena in \ntwo SYM. To study
a model which exhibits a CMS in the weak coupling region, one 
approach is to extend the gauge group to rank greater 
than one\footnote{Alternatively, one can add hypermultiplet matter with
a large mass. In this case there is a discontinuity in 
the spectrum of quark monopole bound states on a CMS curve, which 
has been studied by Henningson \cite{henn}, and the mechanism involves
delocalization in a manner 
analogous to that discussed in this section.}.
Here we consider one of the simplest examples of this kind 
with gauge group SU(3). In the Coulomb phase this
theory exhibits BPS dyon solutions with electric and magnetic charges
associated with either of the unbroken U(1)'s of the Cartan torus.
After choosing a convenient basis of simple roots for the algebra,
one can classify the BPS monopole solutions into one of two types:
those whose magnetic charge is aligned along a simple root --
`fundamental monopoles' -- and those whose magnetic charge is aligned
along the non-simple root. These `composite monopoles' generically
possess CMS curves at weak coupling, and so their dynamics in
this regime is amenable to a semi-classical consideration.

Composite dyons in this, and the closely related
\nfour system, have recently been studied in some 
detail \cite{ly,tong,blly,bly,bl,n2quant,ll}, with the conclusion
that the low energy dynamics of two fundamental dyons 
at generic points of the Coulomb branch acquires 
an additional potential term. This term is associated with the misalignment of
the adjoint Higgs vevs of the two dyons, and leads to the formation of
composite BPS dyons as bound states in this system. We will
review some of these results below, and emphasize the implications
for the dynamics in the near CMS region. The removal of
the composite state on the CMS again arises through
delocalization.

\subsection{The BPS mass formula}

We first review the features of the classical BPS mass formula for \ntwo
SYM with higher rank gauge groups (see e.g. \cite{holl,n2quant}), limiting 
ourselves to SU(N). For the consideration of solitonic
mass bounds, we need consider only the bosonic Hamiltonian which
has the form,
\be
 H =  2{\rm Tr}\int\! {\rm d}^3 x\left\{ \frac 1 2 (E_i)^2+\frac 1 2 (B_i)^2
    +D_0\Ph^*\,D_0\Ph+D_i\Ph^*\,D_i\Ph+\frac 1 2\,[\Ph^*,\Ph]^2\right\},
 \label{Ham}
\ee
where $E_i$ and $B_i$ $(i=1,2,3)$ are the electric and magnetic fields,
and $\Phi$ is the complex adjoint scalar
($\Phi=(\Phi_1+i\Phi_2)/\sqrt{2}$ in terms of the two real adjoint scalars).
We use the
normalization $\Tr \,T^aT^b=(1/2)\delta^{ab}$ for the generators.

The classical vacua satisfy $[\Ph^*,\Ph]=0$, thus requiring
the vev of $\Ph$ to lie in the Cartan subalgebra $ {\bf H}$,
\be
\langle\Phi\rangle=\bm{\phi} \cdot {\bf H}\,.
\ee 
Note that the remaining Weyl freedom
may be fixed by demanding that ${\rm Re}\,\bm{\phi} \cdot \bm{\beta}^a\geq 0$ 
for a given set of simple roots $\{\bm{\beta}^a\}$. This defines a region
which for $|\bm{\ph}\cdot \bm{\beta}^a | \gg \La$
coincides with the semiclassical moduli space of the theory. 
In this region we can safely neglect field-theoretic perturbative and
nonperturbative quantum effects. 
We will only consider the case where the gauge group is maximally broken to
U(1)$^{N-1}$, for gauge group SU($N$). 

For a soliton solution we may define the charge vector $\bm{Q}$
\begin{equation}
\bm{Q}\cdot \bm{H}=(\bm{q}+i\bm{g})\cdot \bm{H}=
\int_{S^2_{\infty}} \! {\rm d}S^i \,( E_i+iB_i )\,,
\end{equation}
where use of the unitary gauge is implied. The real ($\bm{q}$) and 
imaginary ($\bm{g}$) parts of each component of $\bm{Q}$ have 
the interpretation of electric and magnetic charges in the
corresponding U(1); they are are quantized and form a lattice
spanned by simple roots,
\be
 \bm{q} = q_a\, \bm{\beta^a} = e \,n^E_a\,
     \bm{\beta^a}\,,\qquad
\bm{g} = g_a \,\bm{\beta^a} = \frac{4\pi}{e}\,n^M_a \,\bm{\beta^a}\,.
         \label{charges}
\ee
Here $e$ is the gauge coupling, and 
$n^E_a$ and $n^M_a$ are the integral electric and magnetic quantum numbers.
We also normalize the simple roots $\bm{\beta}^a$ with the conventions
$(\bm{\beta}^a)^2=1$, $\bm{\beta}^{a\pm
1}\cdot\bm{\beta}^a=-1/2$, so that the
coroots coincide with the roots.

For general bosonic configurations, there is a Bogomol'nyi 
mass bound following from (\ref{Ham}) which takes the form,
\be
 M \geq {\rm Max}\,\left|{\cal Z}_{\pm}\right| \,,\qquad
 {\cal Z}_{+}  = \sqrt{2}\,\bm{\phi}^*\cdot \bm{Q}\,,\qquad
{\cal Z}_{-}  = \sqrt{2}\,\bm{\phi} \cdot \bm{Q}
\label{ccharge}\,,
\ee
This bound is saturated by solutions of the Bogomol'nyi equations,
\be
 B_i=D_i\, b\,, \;\;\;\;\;\;\;\;\; E_i = D_i\, a\,, \label{BPSeqn}
\ee
along with the equation $D_i^2 a\mp e^2[b,[b,a]]=0$  
which, making use of (\ref{BPSeqn}), 
expresses Gauss' law in the gauge $A_0=\mp a$. 
The fields $a,\,b$  are real and imaginary parts of 
$\exp(i\alpha) \,\Phi$ where the 
angle of rotation $\alpha$ is defined in terms of the charges 
$\bm{Q}$ (see e.g. \cite{fh2,ly,n2quant}).

In the framework of the \nfour supersymmetry algebra 
the parameters $Z_{\pm}$ are
realized as central charges, and it is advantageous to view the system
in this context (implying six instead of two real scalars $\Phi$). 
Within \nfour SUSY it is generally the
case that $|{\cal Z}_+|\neq |{\cal Z}_-|$, and states 
which saturate the Bogomol'nyi
bound $M\geq$ Max $|{\cal Z}_{\pm}|$ will preserve only four of the
sixteen supercharges, and will thus be 1/4 supersymmetric.
If, however, $|{\cal Z}_+|=|{\cal Z}_-|$, states which saturate this
bound will preserve 1/2 of the \nfour supersymmetry. The possibility of having
1/4 BPS states, which only occurs for gauge groups of rank larger than one,
dramatically increases the
number of CMS curves accessible to semiclassical analysis, since
1/4 BPS states generically exhibit regions in the parameter space
where they become marginally stable with respect to ``decay'' into
1/2 BPS states. In this sense it is useful to think of 1/4 BPS
configurations as composite.

The discussion above was framed within \nfour SYM, but this was
simply for orientation. In order to preserve any fraction
of supersymmetry, four of the six real adjoint scalars must vanish
asymptotically, and thus the configurations discussed above all
``descend'' to give classical solutions in \ntwo SYM. The difference
is that now only ${\cal Z}_-$ remains as a central charge \cite{n2quant}
and all states saturating the bound $M=|{\cal Z}_-|$ are 
1/2 BPS states from the point of view of
the \ntwo SUSY algebra. As noted in \cite{n2quant}, those charge
sectors for which $|{\cal Z}_-|< |{\cal Z}_+|$ will have no BPS states
from the point of view of the \ntwo system. Indeed, in this case 
states with $M=|{\cal Z}_-|$ are not  allowed because 
$M\geq |{\cal Z}_+|$. Thus $|{\cal Z}_-|> |{\cal Z}_+|$ is a
necessary condition for the existence of  \ntwo BPS states.

Restricting our attention now to the gauge group SU(3), then
on the Coulomb branch the gauge group is
broken down to the Cartan subalgebra 
U(1)$^2$ and there can be field configurations which are
electrically and magnetically charged under either of these
U(1)'s \cite{weinberg}.  
Following Weinberg \cite{weinberg} we use the term
`fundamental dyons' to refer to those configurations whose charges
are aligned along one of these simple roots. Configurations whose
charges are aligned along non-simple roots
(i.e. $\bm{\beta}^1+\bm{\beta}^2$)
will be referred to as `composite'. We shall focus on a particular
composite configuration which has received considerable attention
in the recent literature -- namely the composite dyon which has equal
magnetic, $n_M^a=(1,1)$, and differing electric,
$n_E^a=(q_1/e,q_2/e)$,  charges along the simple roots.

\subsection{Marginal stability and Coulomb-like interaction}

The BPS mass formula for the (1,1) dyon with $n_M^a=(1,1)$,
$n_E^a=(q_1/e,q_2/e)$  takes the form,
\be
 M_{(1,1)}
=\left|{\cal Z}_{-}\right|
=\sqrt{2}\,\left|(q_{1}+ig)\,\bm{\ph}\cdot\bm{\beta}^1+
            (q_{2}+ig)\,\bm{\ph}\cdot\bm{\beta}^2\right|,
\ee
where $g=4\pi/e$. This configuration has a CMS curve where the (1,1) dyon
is marginally stable with respect to two fundamental dyons:
the first is aligned along $\bm{\beta^1}$, and has  $n_M^a=(1,0)$,  
$n_E^a=(q_1/e,0)$, while the second is aligned along $\bm{\beta^2}$,
and has $n_M^a=(0,1)$, $n_E^a=(0, q_2/e)$. The masses of the fundamental
dyons are 
\begin{equation}
 \label{fuma}
  M_{a} =\sqrt{2}\,
  \left|(q_{a}+ig)\,\bm{\ph}\cdot\bm{\beta}^a\right|
 = m_a \left( 1 + \frac{q_a^2}{2g^2} 
 + {\cal O}\left(\frac{q_a^4}{g^4}\right)\right), \label{massf}
\end{equation}
where in the second equality we have made use of the fact that $e^2\ll
1$ in order to write the dyon mass as a small correction to the 
mass of the corresponding fundamental monopole, 
$m_a=\sqrt{2}g|\bm{\ph}\cdot\bm{\beta}^a|$.
Introducing $\om_a$ as the argument of the vevs,
\begin{equation}
\bm{\ph}\cdot\bm{\beta}^a=\left|\bm{\ph}\cdot\bm{\beta}^a\right|{\rm
e}^{i\om_a}
\label{argu}
\end{equation}
we see that the marginal stability condition $M_{(1,1)}= M_{1}+ M_{2}$
fixes the argument of the ratio 
$\bm{\ph}\cdot\bm{\beta}^1/\bm{\ph}\cdot\bm{\beta}^2$,
\be
\om=\om_1-\om_2\,,
\ee
to be equal to the argument of the ratio of complex charges $Q_2/Q_1$ 
where $Q_i=q_i+ig$. This implies that the CMS equation is $\om=\om_c$ where 
$\om_c$ is defined as
 \begin{equation}
  \label{omegc}
 \sin\om_c=\frac{(q_2-q_1)\,g}
    {\sqrt{g^2+q_1^2}\,\sqrt{g^2+q_2^2}}
\,.
\end{equation}
Provided $n_E^a$ and $n_M^a$ are of a similar order, the angle $\om_c$
is small in the limit $e^2 \ll 1$, i.e. when the electric 
corrections to the dyon
masses are much smaller than the corresponding monopole mass as in 
(\ref{massf}). Thus in this limit the vevs
$\bm{\ph}\cdot\bm{\beta}^a$ are only slightly disaligned, and we can
make use of an expanded version of Eq.~(\ref{omegc}),
\begin{equation}
  \label{omegc1}
 \om_c=\frac{\De q}{g}\left(1+{\cal O}(q_i^2/g^2)\right)\,,\qquad \De q=q_2-q_1
\,. 
\end{equation} 

We are now in a position to verify the general claim of 
Sec.~\ref{domint} that the 
Coulombic interaction vanishes on the CMS. At large distances dyons
can be viewed as point charges which 
interact at rest through electrostatic, magnetostatic, and
scalar exchange. The electrostatic and magnetostatic interactions are
fixed by the corresponding charges, while the scalar exchange
can be read off from the asymptotic form of the Higgs field 
of one of the primary dyons
(in a physical gauge where the configuration
is a linear superposition of the fundamental dyon solutions
\cite{lwy2,fh2}). The effective Coulombic interaction then
takes the form,
\be
 V_{\rm Coul} = -\frac{1}{8\pi r}\left[q_1q_2+ g^2-\sqrt{g^2+q_1^2}
    \sqrt{g^2+q_2^2}\, \cos \om\right]. \label{Veff}
\ee
Similar expressions have appeared in \cite{fh2} and \cite{blly}.
One observes that on the CMS, where the angle
$\om$ is given by Eq.~(\ref{omegc}), the Coulombic potential
$V_{\rm Coul}$ vanishes.

When expanded to second order in $q_i/g$,  
the potential $V_{\rm  Coul}$  takes the form
\be
 V_{\rm Coul}\approx \frac{(\De q)^2 - (g\,\om)^2}{16\pi r}\,, \label{V2}
\label{Vcoul}
\ee
where $\De q=q_2-q_1$ is defined in Eq.~(\ref{omegc1}). 
If the vevs were aligned, i.e. $\om=0$, we
see that to quadratic order, the potential is repulsive
\cite{lwy2,gibbons} (as opposed to the SU(2) case \cite{manton})
and depends only on the electric charge difference $\De q$.
However, for $(g\,\om/\De q)>1$ the potential 
is attractive and the (1,1) BPS dyon 
exists with a mass $M_{(1,1)}$ given in the same approximation by
\begin{equation}
M_{(1,1)}
=\left|{\cal Z}_{-}\right| \approx \left(M_1+M_2\right)
-M_r\,\frac{(\Delta q-g\,\om)^2}{2\,g^2}
\,,\qquad\qquad\left(\frac{g\,\om}{\De q}> 1\right)\,,
\label{zmin}
\end{equation}
where $M_r=m_1m_2/(m_1+m_2)$ is the reduced mass of the monopoles, 
and the corresponding masses $M_{1,2}$ and $m_{1,2}$ are defined in 
Eq.~(\ref{fuma}).
On the other side of the CMS in the range $|g\,\om/\De q|<1$  we have 
repulsion and the the (1,1) BPS dyon does {\em not} exist.

Its interesting to note that
the Coulombic potential reverts to attractive form once more when  
$(g\,\om/\De q)<-1$, and the (1,1) bound states reappear.  
This has a simple interpretation in the framework of 
\mbox{\nfour} supersymmetry:
the lowest mass in this range saturates the $|{\cal Z}_+|$ central
charge (note that now $|{\cal Z}_+|>|{\cal Z}_-|$),
 \begin{equation}
M_{(1,1)}
=\left|{\cal Z}_{+}\right| \approx \left(M_1+M_2\right)
-M_r\,\frac{(\Delta q+g\,\om)^2}{2\,g^2}
\,,\qquad\qquad\left(\frac{g\,\om}{\De q}< - 1\right)\,.
\label{zplus}
\end{equation}
In terms of \mbox{\nfour} SUSY the (1,1) state at $(g\,\om/\De q)<-1$
is 1/4 supersymmetric, but preserves a different subalgebra as compared
to the $(g\,\om/\De q)>1$ case above. Moreover, the generators of
this subalgebra are {\em not} part of the \ntwo superalgebra.
In terms of \ntwo this means that the supermultiplet is not
shortened, but nonetheless the Bogomol'nyi bound is saturated at the classical
level~\footnote{This saturation will be lifted by field-theoretic quantum
corrections.}.  Thus, we see an interesting example where the ``BPS''
nature of the state does not imply multiplet shortening. The presence
of these states in \ntwo SYM was also noted in \cite{n2quant}.

We now address the question of what happens to the
(1,1) state on the CMS, i.e. when $|g\,\om/\De
q|=1$ and the $1/r$ terms in the potential vanish? As we will see in
Sec.~\ref{sec:msd} the dynamics on the CMS is governed by repulsive 
$1/r^2$ terms demonstrating, even at the classical level,
that there is no localized bound state on the CMS.

\subsection{Zero Modes and Moduli Spaces}

We will shortly consider the low energy dynamics of the
fundamental dyons comprising the (1,1) system. 
However, we first recall a few details regarding 
the zero mode structure of dyon
solutions in \ntwo SYM. For BPS dyons in pure \ntwo SYM the 
unbroken \none supersymmetry is enough in this case
to pair the bosonic and fermionic zero modes \cite{gauntlett} so we shall
focus here just on the bosonic modes.
Generic dyon solutions, corresponding to the
embedding of the SU(2) monopole along some root of SU(3) have
four bosonic zero modes \cite{weinberg,holl} parametrizing the moduli space,
\be
 {\cal M}_1 = {\bf R}^3 \times S^1.
\ee
These modes are naturally identified as the center of mass position
in ${\bf R}^3$ and the $S^1$ is an isometry conjugate to the
conserved electric charge.

For dyons embedded along a simple root, this is the moduli space
for all choices of field-theoretic moduli. However, if we consider
composite monopoles, then the monopole moduli space ${\cal M}$
enlarges to a space of
dimension eight, as is compatible with separating the constituents
into two isolated fundamental monopoles \cite{holl,fh2}. This result
was obtained in \cite{holl} using the index calculations of
Weinberg \cite{weinberg} for real Higgs fields. 

For the case at hand,
the magnetic charge is $\bm{g}=g(\bm{\beta}^1+\bm{\beta}^2)$ and
asymptotically the eight dimensional moduli space 
is simply ${\cal M}_1 \times {\cal M}_1$. However, its exact
form has also been deduced in
\cite{connell,gl,lwy,lwy2},
\be
 {\cal M}_2 = {\rm \bf R}^3 \times \frac{{\rm \bf R}\times {\cal
      M}_{TN}}{Z}.
\ee
The first ${\rm \bf R}^3 $ factor corresponds 
to the center of mass position, while
the second ${\rm \bf R}$ factor refers to the coordinate
conjugate to the total electric charge. The corresponding metric is flat.
The relative moduli space ${\cal M}_{TN}$ is positive mass Taub-NUT
space (which is asymptotically ${\bf R}^3\times S^1$). Its 
four coordinates $z^\mu$ describe the relative distance $r$ between
the cores, with the corresponding polar and azimuthal angles $\th$ and
$\varphi$, and also the relative phase $\ch$, conjugate to the 
relative electric charge $\De q$. The factor $Z$ denotes a 
discrete identification for the charge coordinates, ensuring
that the asymptotic geometry has a compact factor $S^1\times S^1$, associated
with the conserved charges.

The Taub-NUT metric, in our conventions~\footnote{We follow \cite{n2quant}
with the exception that $\chi$ is rescaled to have a period of $2\pi$
rather than $4\pi$, and consequently the conjugate momentum is
integer ($(q_2-q_1)/e$) rather than half-integer ($(q_2-q_1)/2e$) valued.} 
takes the form,
\be
\label{gtn}
 ds^2_{\rm TN} = g^{\rm TN}_{\mu\nu}\,dz^\mu dz^\nu = m(r)\! \left[
d r^2 + r^2 d\theta^2 +r^2 \sin^2\!\theta\, d\varphi^2\right]  +
\frac{g^2}{e^2\,m(r)}
        \left[d\ch + \frac 1 2\cos \theta\, d\varphi\right]^2,
\ee
with a `running' mass parameter,
\be 
 m(r) = M_r + \frac{2\pi}{e^2 r},
\ee
which asymptotes to the reduced mass $M_r$
when the relative separation $r$ diverges. 

In terms of the internal U(1) angles $\ps_i$ of the 
fundamental monopoles, the combination $\xi=\ps_1+\ps_2$ is conjugate
to the total electric charge $q_t$ (or more precisely, to $q_t/e$),
\be
 q_t = \frac{m_1q_1 + m_2 q_2}{m_1+m_2}\,, 
\ee
while
$\ch=(m_1\ps_2-m_2\ps_1)/(m_1+m_2)$ is conjugate to $\De q/e$, i.e. to 
the relative electric charge introduced above \cite{gl,lwy}.

\subsection{Moduli space dynamics}
\label{sec:msd}

As first discussed in this context by Manton \cite{mantongeo}, 
the low-energy dynamics of fundamental monopoles may be understood 
as geodesic motion on the underlying moduli space. This picture
extends to dyon solutions with aligned charges, but recent work
on the dynamics of the fundamental constituents of the
(1,1) system \cite{ly,bhlms,tong,blly,bly,bl,n2quant}
has shown that for two fundamental dyons with misaligned
charges the Lagrangian following from the
geodesic approximation needs to be corrected by a new 
term \cite{blly,n2quant}.  In this subsection we will partially review
these results, while emphasizing the features of the
near CMS region.

The construction of Refs.~\cite{blly,n2quant} can be reformulated in
terms of the following Lagrangian for the relative moduli 
$z^\mu=\{{\vec r},\chi\}$
\be
 L_{\rm rel} = \frac{1}{2}\,g^{\rm TN}_{\mu\nu}\, \dot{z}^{\mu}\,\dot{z}^{\nu}
          + g^{\rm TN}_{\mu\nu}\,  \dot{z}^{\mu}\, G^{\nu}\,, 
\label{Lag}
\ee
where  the metric $g^{TN}_{\mu\nu}$ is given by Eq.~(\ref{gtn}) and
the `potential' $G^{\nu}$, which is a Killing vector generating the 
$\chi$ isometry, is
\be
G^{\nu}=\frac{e}{g}\,M_r\,\om\,\delta^\nu_\chi\,.
\ee
Here $\om$ is the angle of disalignment between
the condensates $\bm{\ph}\cdot\bm{\beta}^a$, see Eq.~(\ref{argu}).
The term containing $G^{\mu}$ is dynamically significant due to 
nontrivial fibering of the $S^1$ associated with $\chi$ 
in the Taub-NUT metric.  

The classical Hamiltonian then has the form
\begin{equation}
H_{\rm rel}=\frac{1}{2}\, g_{\rm TN}^{\mu\nu} \,
    (\pi_{\mu}-G_{\mu})(\pi_{\nu}-G_{\nu})
\,, \label{Hk}
\end{equation}
where $\pi_{\mu}=g^{\rm TN}_{\mu\nu}({\dot z}^\nu +G^\nu) $ are the canonical
momenta. In terms of the original field theory this Hamiltonian is
interpreted as $M-|{\cal Z}_-|$, and thus BPS states 
are `vacuum' states of this Hamiltonian.

Two of the canonical  momenta, namely $\pi_\varphi$ and $\pi_\chi$,
are conserved quantities conjugate to isometries along the azimuthal angle
$\varphi$ and the phase $\chi$. The value of  $\pi_\varphi$ is 
the $z$-projection of the angular momentum $l_z$, and 
$\pi_\chi$ is equal to $\De q/e$.  Substituting
these and the inverse metric $g_{\rm TN}^{\mu\nu}$ into Eq.~(\ref{Hk})
we obtain
\begin{equation}
H_{\rm rel}=\frac{\pi_r^2}{2\,m(r)} +\frac{\pi_\theta^2}{2\,m(r)\,r^2} +
\frac{1}{2\,m(r)\,r^2\sin^2\!\theta}\left(l_z-\frac{\De q}{2e}\cos \theta\right)^2
+ V(r)
\,,\label{Hk1}
\end{equation}
where
\begin{equation}
V(r)=\frac{M_r}{2\,g^2}\left(1+\frac{2\pi}{e^2\,M_r\,r}\right)^{-1}
\left(\De q - g\,\om +\De q \,\frac{2\pi}{e^2\,M_r\,r}\right)^2
  \label{Vr}
\end{equation}
is the only term in $H_{\rm rel}$ which depends on the field-theoretic
moduli $\bm{\phi}$ (via $\om$).

This Hamiltonian vanishes at $\pi_r=0$, $\pi_\theta=0$, and the equilibrium
values $r_0$ and $\theta_0$ of the corresponding coordinates are given by
\begin{equation}
 r_0 = \frac{2\pi}{e^2\,M_r} \,\frac{\De q}{g\,\om -\De q}\,,
\qquad \cos \theta_0=2\,l_z\,\frac{e}{\De q}\,,
\label{r0}
\end{equation}
when $(g\,\om/\De q)>1$ and $|2l_z\,e/\De q|<1$. There is no solution 
for $(g\,\om /\De q)<1$, i.e. the BPS state ceases to exist upon 
crossing the CMS where $g\,\om =\De q$. 
We see that the system describes the composite state as a 
bound state of the dynamics  whose spatial size, corresponding
to the separation of the primary constituents, diverges on approach
to the CMS. 

It is instructive to 
expand the potential $V(r)$ at large $r$ 
\be
V(r)=M_r\,\frac{(\De q - g\,\om)^2}{2\, g^2}+
\frac{(\De q)^2 -(g\,\om)^2}{16\pi\,r}+
\frac{(g\,\om)^2}{8\,e^2\,M_r\,r^2}+\ldots\,,
\ee
where we omitted $1/r^3$ terms and higher powers of $1/r$. The constant 
term $V(r\to \infty)$ marks the start of the continuum. Indeed, 
adding $|{\cal Z}_-|$ from Eq.~(\ref{zmin})
we obtain $M_1+M_2$ in the limit $r\to\infty$. The $1/r$ term
coincides with the Coulombic potential
(\ref{Vcoul}) discussed earlier.  It provides attraction 
for $|g\,\om/\De q|>1$, the range where the bound states exist. The range 
$(g\,\om/\De q)<-1$ corresponds, as discussed above, 
to $M_{(1,1)}=|{\cal Z}_+|$. What we see in addition is 
the repulsive $1/r^2$ term which leads to the existence of
an equilibrium at large $r_0$ near the CMS. However, on
the CMS it becomes the dominant term, and 
so there is  no localized state on the CMS.

We conclude this section with some brief comments about quantization.
The quantization of the dyon system in the context of \ntwo SYM was
first discussed in detail by Gauntlett \cite{gauntlett} in the case
where the Higgs vevs are aligned, and this discussion has since been
generalized to the case considered here by Bak et al. \cite{bly} for
\nfour, and by Gauntlett et al. \cite{n2quant} for \ntwo SYM. The crucial
feature of this system is that the relative moduli space inherits
a triplet of complex structures $J^{(a)}$, $a=1..3$, and
is a hyperk\"ahler manifold. Consequently, the system exhibits \nfour
supersymmetric quantum mechanics with four real supercharges $Q_A$
where the index $A$ can be associated loosely with a quaternionic 
structure $J^{(A)}=(\bm{1},J^{a})$. These supercharges satisfy the
superalgebra,
\be
 \{Q_A,Q_B\} = 2 \de_{AB} H_{\rm rel},
\ee
where $H_{\rm rel}$ is the supersymmetric completion of the Hamiltonian
defined in (\ref{Hk}), to be interpreted as $M-|{\cal Z}_-|$. One can
compare this with the SQM constructed in Sec.~2.

An interesting feature of this system is that the symmetries of the 
superalgebra and the moduli space combine to ensure that the wavefunctions
have a nontrivial dependence on the angular moduli, as well as the
relative separation $r$. Specifically,
the ground state wavefunction has the functional form \cite{bly}
$\Ps_0=\Ps_0(r,\si_a)$ where $\si_a$ are the basis 1-forms on the $S^3$
parametrized by $(\th,\varphi,\ch)$. This dependence is hinted at through the
$\th$-dependent terms in the Hamiltonian (\ref{Hk1})  above. 
One may speculate that because of the high degree of symmetry in this system
-- the bosonic system possesses an additional conserved quantity 
of Runge-Lenz type \cite{ll} -- a more precise separation of variables
may be possible, but we will not pursue this issue here. We note only
that, as demonstrated above, the
delocalization on the CMS is associated with the cancellation of the 
terms of ${\cal O}(1/r)$ and depends purely on the relative
separation. Moreover, this conclusion in the
classical bosonic system apparently extends to SQM \cite{bly}.

\section{Delocalization via Massless Fields}
\label{sec:six}

On particular submanifolds of the CMS, the discussion we have presented
above may be incomplete because certain fields may become
massless. Indeed, generically there will be particular points on the CMS where
states which are stable on both sides are massless. The presence of these
singularities in moduli space can then be thought of as the `origin'
of the CMS since marginally stable states may not be single valued
on traversing a contour around the singularity, and so a discontinuity
in the spectrum becomes necessary for consistency. For example,
this point of view provided one of the
first arguments for the fact that the W boson must be removed from the
spectrum inside the strong coupling CMS in \ntwo SYM \cite{n2cms}.

In this section we will discuss the behavior of BPS states near
these singular points. Within a simple 2D Wess-Zumino model 
we will find that the
discontinuity of the BPS spectrum is explained by the delocalization
of fermionic zero modes of the soliton on the CMS. The CMS in this
case corresponds to the `collision' of two vacua in the parameter
space, and thus one might anticipate similar phenomena in \ntwo SYM
near Argyres-Doulgas points \cite{ad} when the singularities
associated with monopole and quark vacua collide. Unfortunately, this
occurs at strong coupling and is out of the range of our
semi-classical analysis.

\subsection{Breaking ${\cal N}=1$ to ${\cal N}=1$
 and  the restructuring of WZ solitons}

The model is a simplified version of the \ntwo  Wess-Zumino model \cite{WZ}
in two dimensions considered in Sec.~\ref{sec:three}. We shall set the second
field $X$ to zero, but consider a new perturbation which breaks
\ntwo down to \none$\!$. This setup was introduced in Ref.~\cite{svv} 
(see Sec.~8).

The superpotential prior to perturbation is then of the standard
Landau-Ginzburg form, and its worthwhile recalling a few pertinent
details of these theories. A general classification of 
the \ntwo  Landau-Ginzburg--type theories
in two dimensions was given in \cite{Vafa}, while construction of the
representations of the ${\cal N}\!\!=\!2$ superalgebra
with central charges was presented in Refs.~\cite{Ken}.
It was shown that the supermultiplet of BPS soliton states is
shortened, and  this shortened multiplet consists of two states
$\{u,~d\}$ as we discussed earlier in Sec.~\ref{sec:three}. In particular, in 
${\cal N}\!\!=\!2$ theories there exists a conserved fermion charge
$f$. The fermion charge of the $u$ and $d$  states is fractional but 
the difference is unity, $f_u -f_d=1$.   

What changes on passing to ${\cal N}\!\!=\!1$ in two dimensions?
The irreducible representation of the ${\cal N}\!\!=\!1$ algebra for
BPS states is now one-dimensional (to the best of our knowledge this
was first noted in  Ref.~\cite{svv}).
The only remnant of the fermion charge is a discrete subgroup $Z_2$
which is spontaneously broken.

It is natural then to expect a restructuring of the BPS spectrum
when ${\cal N}\!\!=\!2$  is broken down to ${\cal N}\!\!=\!1$. 
 We will study the manner in which restructuring occurs by
considering the spectrum of fermionic zero modes of a soliton solution
as we vary the soft breaking parameter $\mu$. We observe that for
small $\mu$ the BPS spectrum remains the same as in the
unbroken ${\cal N}\!\!=\!2$ theory. 
However, starting at a critical value $\mu_*$ -- corresponding to
a `point of marginal stability' -- half the BPS states 
disappear from the spectrum.  This occurs because
quasiclassically the counting of states in the supermultiplet is
related to the counting of zero modes of the
soliton and when $\mu$ reaches $\mu_*$ some of the fermionic zero modes
become non-normalizable.  To follow their fate one can 
introduce a large box. Then the number of states does not change, but
at $\mu=\mu_*$ the identification of states with zero modes
implies that half the BPS states spread out all over the box while
for $\mu>\mu_*$ they lie on the boundary of the box and are removed
from the physical Hilbert space.  
This picture is quite analogous to the quantum mechanical discussion
in Sec.~\ref{sec:two}. However, as we shall see, the quantum description is
complicated here by the presence of a massless field.

We take the K\"ahler metric to be canonical and the
cubic superpotential of the model is
conveniently represented in terms of real bosonic variables
$\varphi_i$, $i=1,2$,
\begin{equation}
{\cal W}(\varphi_1, \varphi_2) = \frac{m^2}{4\lambda} \,\varphi_1 -
\frac{\lambda}{3} \,\varphi_1^3 +\lambda\,\varphi_1\varphi_2^2~.
\end{equation}
In fact, ${\cal W}(\varphi_1, \varphi_2) $ is harmonic
\begin{equation}
\frac{\partial^2 {\cal W}}{\partial \varphi_i \partial
\varphi_i}=0\qquad \mbox{for ${\cal N}=2$}
\,.
\end{equation}
as it is the imaginary part of the four dimensional superpotential which is
analytic in $\varphi_1 + i \varphi_2$. This is therefore a reflection
of \ntwo supersymmetry in two dimensions.

Now, to break ${\cal N}\!\!=\!2$ down to ${\cal N}\!\!=\!1$
we consider a more general, nonharmonic,
superpotential ${\cal W}\,(\varphi_1, \varphi_2) $,
\begin{equation}
{\cal W}\,(\varphi_1, \varphi_2) = \frac{m^2}{4\lambda} \,\varphi_1
-
\frac{\lambda}{3} \,\varphi_1^3 +\lambda\,\varphi_1\varphi_2^2
+\frac{\mu}{2}\,\varphi_2^2\,,
\end{equation}
where $\mu$ is the soft breaking parameter.
There are two vacuum branches,
\be
 \left\{\varphi_1=\pm\frac{m}{2\la};~~~\varphi_2=0\right\},\;\;\;\;\;\;
 \left\{\varphi_1=-\frac{\mu}{2\la};
 ~~~\varphi_2=\pm\frac{\sqrt{\mu^2-m^2}}{2\la}\right\},
\ee
but the second exists only for $\mu>m$, and vacua collide when
$\mu=m$.

This model exhibits a classical kink solution which interpolates
between the first set of vacua, and is given by
\be
 \varphi_1 = \frac{m}{2\la}\tanh \frac {m z}{2},\;\;\;\;\;\varphi_2=0.
\label{soliton}
\ee
It satisfies the classical BPS equations,
\beq
\frac{\partial \varphi_i}{\partial z} = \frac{\partial {\cal
W}}{\partial
\varphi_i}\,.
\eeq
The zero modes corresponding to this kink are as follows:

\noindent
(a) one bosonic mode:
\begin{equation}
\chi_0 = C\, \frac{1}{\cosh^2 (mz/2)}
\label{chi}
\end{equation}
in the field  $\varphi_1$ corresponds to
(the spontaneous breaking of) translational invariance,
$$
\chi_0 \propto
{\rm d}
\varphi_1/ {\rm d} z\,.$$
The constant $C$ in Eq. (\ref{chi}) is a normalization constant; its
explicit
numerical value is not important. (Below
the normalization constants in the zero modes will be omitted.)

\noindent
(b) two fermionic modes:   The first zero mode
in the field $\psi_{1,2}$ (the indices number the superfields and
fermionic
components in the basis where $\gamma^0=\sigma_2$,
$\gamma^1=i\sigma_3$)
has the same form $\chi_0$ as the translational mode. It is not accidental,
the corresponding differential operators are the same  due to
\none supersymmetry. The second
fermionic zero
mode
\begin{equation}
\xi_0 = \frac{\exp (-\mu z)}{\cosh^2 (mz/2)}
\label{eta}
\end{equation}
appears in the field $\psi_{2,1}$. At $\mu=0$ the existence of this
mode
is a
consequence of the ${\cal N}\!\!=\!2$ SUSY, and  $\xi_0$ coincides
with
$\chi_0$.
At nonvanishing $\mu$,
when the extended SUSY is broken,  this zero mode is maintained
 by virtue of the Jackiw-Rebbi index
theorem \cite{jr}.

\begin{figure}[ht]
\epsfxsize=9cm
 \centerline{%
   \epsfbox{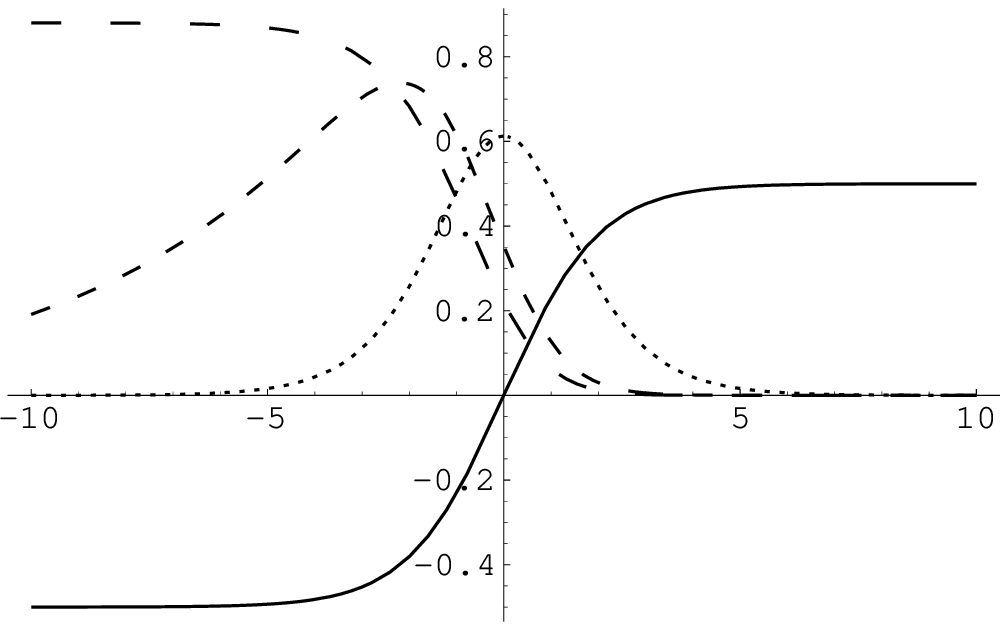}%
         }
 \newcaption{The kink profile (solid line), with the
zero mode $\xi_0$ for  $\mu/m=0$ (dotted line), $\mu/m=0.8$
   (short-dashed line), and $\mu/m=1$ (long-dashed line). Note that
   the vertical scale has been altered for ease of
presentation.\label{fig:xiall}}
\end{figure}

An interesting feature of the zero mode $\xi_0$ is that it is
asymmetric in $z$
for $\mu \neq 0$. Moreover, this mode is normalizable only for
\begin{equation}
\mu < m\,.
\end{equation}
This is readily seen from its asymptotics,
\begin{equation}
\xi_0(z\rightarrow \infty) \sim e^{-(\mu+m)z}\,, \qquad
\xi_0(z\rightarrow -\infty) \sim e^{-(\mu-m)z},
\end{equation}
The explicit form of the zero mode for few values of $\mu$ is
exemplified in Fig.~\ref{fig:xiall}.
The loss of normalizability occurs at $\mu=m$, when
$$
{\rm det}\,  \left\{\frac{\partial^2 {\cal
W}}{\partial
\varphi_i
\partial \varphi_j}\right\} = 0
$$
in one of the vacua between which the soliton solution interpolates.
In other words, one of the vacuum states has gapless excitations at this
point. Indeed, in the $z\rightarrow -\infty$ vacuum, the eigenvalues of
the fermion mass matrix are: $m$ and $\mu-m$, and thus indeed at the
point $\mu=m$ where the vacuum branches meet, there is a massless
field. This system represents a simplified analog of an Argyres-Douglas
point in that the massless field arises through the collision of vacua.
Furthermore, we see that this infrared effect destabilizes one of the
fermionic zero modes of the soliton.

To study the infrared behavior in detail let us put the system in a
large box, i.e. impose boundary conditions at $z=\pm L/2$
where $L$ is large but finite. We choose these
conditions in a form  which preserves the remnant
supersymmetry in the soliton background (i.e. the BPS nature of the
soliton),
\begin{eqnarray}
&&\left. \left(\partial_z\varphi_i - \frac{\partial {\cal W}}{\partial
\varphi_i}
\right)
\right|_{z=\pm L/2}\!\!\!=0\,,
\nonumber\\[0.2cm]
&&\left. \left(\delta_{ij}\partial_z - \frac{\partial^2 {\cal
W}}{\partial
\varphi_i
\partial \varphi_j}\right)
\psi_{j2}\right|_{z=\pm  L/2}\!\!\!=0 \,,\quad
\psi_{i\,1}{\Big|}_{z=\pm  L/2}\!\!=0\,,
\label{boundary}
\end{eqnarray}
(see Ref.~\cite{svv} for details).
It is easy to check that the soliton solution (\ref{soliton}) as well as
the zero
modes (\ref{chi}), (\ref{eta}) are not deformed by these boundary
conditions, i.e. (\ref{soliton}) remains a solution of the classical
BPS equations with the appropriate boundary conditions at finite $L$.

In the finite box there is no problem with normalization;
 the zero mode  (\ref{eta})
remains a solution of the Dirac equation in the soliton background
for all $\mu$.
However, at $\mu>m$ the mode is localized on the left wall of the
box
instead of
sitting on the soliton as  is the case at $\mu<m$. Thus, at $\mu=m$ the
critical
phenomenon
of delocalization starts. As we will show below, upon
quantization
this means
that some BPS soliton states have disappeared 
from the physical Hilbert space.

\subsection{Quantization}

We shall not present a detailed analysis of the quantization of the
system here as it requires a somewhat different treatment to the
supersymmetric quantum mechanics we have considered thus far. In this
case, one needs to consider the dynamics of the light field
in addition to the collective coordinates of the soliton.

However, provided we only consider a region somewhat away from the
CMS, the spectrum is easily determined. As usual, the remnant
\none supersymmetry pairs the nonzero modes (one bosonic to two
fermionic) around the soliton (see e.g. Sec.~3G of \cite{svv}),
and the relevant contributions cancel.
Thus the soliton spectrum is determined by the zero modes,
corresponding to which
we have one bosonic collective coordinate $z_0$ corresponding to the
center of the kink, and two real Grassmann collective coordinates
$\al_1$ and $\al_2$ determined by the zero modes,
$\ps_{1,2}=\al_1\ch_0(z)+\cdots$
and $\ps_{2,1}=\al_2\xi_0(z)+\cdots$. Combining them into one
complex parameter $\et=(\al_1+i\al_2)/\sqrt{2}$, the collective
coordinate dynamics at $\mu=0$ is determined by the quantum mechanical
system
\be
 {\cal L}_{eff} = -M + \frac{1}{2}M\dot{z}_{0}^2+{iM}\bar\eta\dot\eta,
\ee
where $M$ is the physical kink mass, which we can set to one.

The quantization is carried out in the standard manner.
If $z_0$ and $\eta$ are the canonical coordinates,
we introduce the canonical momenta
\beq
\pi_{z_0} = \dot{z}_0\,,\quad \pi_\eta = -i\bar\eta\,,
\label{canmom}
\eeq
and impose the (anti)commutation relations
\beq
[\pi_{z_0}, z_0] = -i\,,\quad \{\pi_\eta , \eta\} = -i\,.
\label{comrel}
\eeq
One then proceeds to construct the raising and lowering operators
in the standard manner. From Eqs. (\ref{canmom}) and (\ref{comrel})
it is clear that $\eta$ can be viewed as the lowering operator, while
$\bar\eta$ is the raising operator. One then defines the `vacuum state'
in the kink sector by the condition that it is annihilated
by $\eta$,
$$
\eta |`{\rm vac }'\rangle = 0\,.
$$
The application of $\bar\eta$
produces a state which is degenerate with the vacuum state.
$|`{\rm vac }'\rangle$ and $\bar\eta |`{\rm vac }'\rangle$ are two
quantum states which form a (shortened)
representation of \ntwo supersymmetry (at $\mu =0$).
It is clear that these two states, which are degenerate in mass,
have fermion numbers differing by unity.

What happens at $\mu\neq 0$? At $\mu < m$ the situation is exactly
the same as at $\mu =0$ (apart from the fact that
the fermion number is not conserved now
and we must classify the states  with respect to their
$Z_2$ properties). We have two degenerate quantum states,
both are spatially localized and belong to the physical sector
of the Hilbert space. At $\mu = m$ only the vacuum state
is localized. The spatial structure of the second state $\bar\eta
|`{\rm vac }'\rangle$ has a  flat component,
which extends to the boundaries of the box.
At $\mu > m$ this component is peaked at the boundary.
The easiest way to see this is to introduce an external source
coupled to $\varepsilon_{ij}\bar\psi_i\psi_j$.
Thus, the state $\bar\eta |`{\rm vac }'\rangle$
disappears from the physical sector
of the Hilbert space. The supercharge $Q_1$ acting on the state
$|`{\rm vac }'\rangle$ produces this state itself, rather than
another state. (We recall that $Q_2$ annihilates $|`{\rm vac
}'\rangle$.) Formally this looks like a spontaneous breaking of the
remnant supersymmetry.

\section{Concluding Remarks}
\label{sec:seven}

Using quasiclassical methods we have argued that the underlying
dynamics of marginally stable solitons is determined (generically)
by non-relativistic supersymmetric quantum mechanics. Composite BPS states 
which disappear on the CMS were found to do so through a process
of delocalization in coordinate space. Within the quantum mechanical
description this process was associated with the bound state level
reaching the continuum, while further progress beyond the CMS
leads to a potential with a non-supersymmetric ground state.
This is a generic picture. In certain cases the CMS can be a boundary between
sectors with different composite solitons, the quantum mechanical potential 
then vanishes on the CMS. 

One of the crucial features allowing a detailed investigation of the
effective quantum mechanical dynamics in the two field model
considered in Sections \ref{sec:three} and \ref{sec:four} was the
linear realization
of supersymmetry in the two-soliton sector of the
non-relativistic SQM system. This embedding in 1+1D is similar to 
the 3+1 D effective dynamics of two BPS dyons in \ntwo and ${\cal N}\!\!=\!4$ 
supersymmetric gauge theories \cite{blly,bly,bl,n2quant}. 
In 3+1D the effective Coulombic interaction governs dynamics near the CMS.
We demonstrated this in Section \ref{sec:five} for the composite dyon in
SU(3), showing how the BPS states  swell upon the approach to the CMS.

One may wonder whether some of the conclusions noted above
regarding the behavior of composite bound states near the CMS
might not be artifacts of the quasiclassical approximation. In particular,
returning to pure \ntwo SYM with gauge group SU(2),
the classic example of marginal stability with which we started
this discussion is that of the $W$ bosons on a CMS curve at strong
coupling \cite{fb1,bf2}, for which our methods are not directly
applicable. We are going to dwell on this issue in a separate 
publication \cite{RV}.
Here we will briefly present two suggestive arguments pointing to the
conclusion that this phenomenon involves delocalization in 
the same manner as the examples we have discussed.

The first observation involves duality. If we consider a region very
close to the CMS for the $W$ boson and not too far from the monopole
singularity, we can consider a point particle approximation for the 
monopoles and dyons within the dual magnetic description.
Provided we are close enough to the CMS, a nonrelativisitic
approximation will be reliable, and from this viewpoint the 
dissociation of the $W$ is superficially quite similar to that of 
the bound states of dyons discussed in Sec.~\ref{sec:five}, with the 
roles of electric and magnetic charge reversed.

The second observation involves the realization of the BPS states
considered here in terms of string junctions \cite{bergman,ly,ohtake,Wsj}
in Type IIB string theory and its extension to F-theory \cite{sen}.
Although there are still subtleties with this realization, 
specifically concerning a mismatch between field- and string-theoretic
counting of bosonic moduli \cite{bhlms}, its interesting that 
the disappearance of marginally stable states in this framework appears
to universally implies delocalization. The crucial point is that
this process involves shrinking one or more of the spokes of the 
junction to zero length, while it has recently been pointed out
\cite{kk} that the equilibrium separation of the two constituent 
states in the field theory is inversely proportional to the 
length of the shrinking prongs.

As a final remark, its worth commenting on additional subtleties which
arise when considering extended BPS objects. In particular, although
we concentrated here on BPS particles, the notion of
marginal stability is more general as supersymmetry algebras
may also admit central charges supported by extended BPS objects such
as strings and domain walls. Indeed, our
classical analysis in Sec.~\ref{sec:three} may be lifted to four
dimensions where the kink solutions describe BPS domain
walls. However, we concentrated on particle states specifically for
the reason that quantization in this case leads to quantum mechanics, which is
of course well-understood. The main technical difficulty in extending
these arguments to solitons such as domain walls is that in addition to the
dynamics of relative collective coordinates, one also needs to
consider the massless sector of the field theory on the worldvolume of
the soliton.

\bigskip
\section*{\bf Acknowledgements}
\addcontentsline{toc}{section}{\numberline{}Acknowledgments}

We are grateful to A.~Gorsky, A.~Losev, A.~Morozov, N.~Nekrasov, A.~Rosly,
V.~Rubakov,  A.~Smilga, M.~Strassler, D.~Tong,
and A.~Yung for useful discussions and comments. AR would like to 
thank the Theoretical Physics
Institute at the University of Minnesota, where part of this 
work was carried out. This work
was supported by  DOE under the grant number
DE-FG02-94ER408.

\section*{Appendix: Classical soliton potential}
\label{sec:append}

\addcontentsline{toc}{section}{\numberline{}Appendix: Classical
soliton potential}
\renewcommand{\theequation}{A.\arabic{equation}}
\setcounter{equation}{0}

In this appendix we present an alternative derivation of the
classical potential $(W'(s))^2$ entering the SQM Hamiltonian. This
approach is purely bosonic, but requires knowledge of the composite
soliton solutions obtained in Sec.~\ref{sec:three}.

We start from the expression (\ref{deltae}) for the binding energy
$\Delta E_+$,
\beq
\Delta E_+ =  - {m^3 \over \lambda^2} \, {3 \, (\nu_2-\mu_2 \, \kappa)^2
\over \kappa \, \nu_{1+}\nu_{1-} }~.
\label{deltaer}
\eeq
(For $\nu_2/\mu_2 < 0$ the relative sign between $\nu_2$ and $\mu_2$ in
this expression must be reversed.)
The formula (\ref{deltaer}) gives the minimum of  $U(s)-U_+$,
where $U_+$ is the value of potential at $s=\infty$,
while the position of the minimum in $s$ is given by
Eq.~(\ref{condition}). 
We can combine these two results in order to find the
classical expression for $U(s)$ by using the standard Legendre
transform approach.  We introduce a source term
in the original superpotential (\ref{wphix}), thus replacing ${\cal
W}(\Phi, \, X)$ by
\beq
{\tilde {\cal W}}(\Phi,\, X; \, j)={\cal W}(\Phi, \, X) - {m^2 \over
\lambda} \, j \, X ~,
\label{source}
\eeq
where $j$ is a dimensionless (and in general complex) parameter 
corresponding to the strength of the source.
For a static classical configuration described by this superpotential
the calculation of the energy in fact gives the minimum of the functional
${\cal E}(j)$:
\beq
{\cal E}(j)=E(j) + \left ( j \, {m^2 \over \lambda} \, \frac 1 2
\int{\rm d} z \,{\rm
d}^2\theta\,X(x, \theta) +{\rm h.c.} \right )~,
\label{pseudoen}
\eeq
where $E(j)$ is the value of the original energy on the configuration
which extremizes the action at a given source strength $j$, and $X(x,
\theta)$ is the $X$ superfield evaluated on that configuration. 
(In fact, being static, $X$ does not depend on time.)

Clearly the effect of the source term is equivalent
to a shift in $\nu$: $\nu \to \nu + j$ and for our
purposes it is sufficient to consider a purely imaginary source, $j=
{\rm i} \, j_2$. Then the $s$ dependent part of the functional ${\cal
E}(j)$ for the two-soliton static configuration is read directly from equation
(\ref{deltaer}) after replacing $\nu_2$ by $\nu_2+j$:
\beq
\Delta {\cal E}(j_2)= - {m^3 \over \lambda^2} \, {3 \, (\nu_2 + j_2
-\mu_2 \, \kappa)^2 \over \kappa \, \nu_{1+}\nu_{1-} }~,
\label{deltaper}
\eeq
and the relation between the $j$ dependent equilibrium position $s$ and
the value of $j$ is derived from Eq.~(\ref{condition})
\beq
\nu_2 + j_2=  \mu_2 \, \kappa \, w_{\nu_1}(m \, s)~.
\label{newcond}
\eeq

The quantity of interest for us here, however, is not the functional
${\cal E}(j)$ as a function of $j$, but rather the binding energy
$\Delta E$ as a function of $s$. The latter is found in the standard way
from the relation
\beq
\Delta E = {\cal E}(j) - j { {\rm d} {\cal E} \over {\rm d} j}
\eeq
with the variable $j$ being eliminated in favor of $s$, using the
relation (\ref{newcond}). Performing this simple operation on the
expressions (\ref{deltaper}) and (\ref{newcond}) one finds
\beq
\Delta E (s) = U(s)-U_+={m^3 \over \lambda^2} \, {3 \, (\nu_2  -\mu_2 \,
\kappa
\, w_{\nu_1}(m \, s))^2 \over \kappa \, \nu_{1+}\nu_{1-} } - {m^3 \over
\lambda^2} \, {3 \, (\nu_2  -\mu_2 \, \kappa)^2 \over \kappa \,
\nu_{1+}\nu_{1-} } ~,
\label{deltaeofs}
\eeq
which represents the classical interaction energy of two
primary solitons separated by a distance $s$. Naturally, the minimum of
$\Delta E(s)$, as found from this expression, coincides with that given
by equation (\ref{deltaer}) at the separation $s$ determined by
(\ref{condition}).

Comparing the last term on the right hand side of
Eq.~(\ref{deltaeofs}) with the
expression (\ref{upm}) for $U_+$ we see that they coincide. Thus,
the  potential is
\beq
U_{\rm cl} (s) = {m^3 \over \lambda^2} \, {3 \, (\nu_2  -\mu_2 \, \kappa
\,
w_{\nu_1}(m \, s))^2 \over \kappa \, \nu_{1+}\nu_{1-} }~,
\label{potofs}
\eeq
where the subscript reminds us that this is the potential found at the
classical level. This result
correctly reproduces the energy difference between the
asymptotic states at both infinities. With this normalization,
one may readily check that the
the minimum (zero) of the potential corresponds in (\ref{deltae}) to the
mass of the BPS $\{1,1\}$ soliton,
\beq
M_{1,1}= {8 \over 3} \, {m^3 \over \lambda^2} \kappa +{3 \over 4} \,
{m^3 \over \lambda^2} \, {(\nu_{1+}-\nu_{1-})^2 \over
\kappa}~.
\label{bps11}
\eeq

Comparing the above expression for the potential with the general form
of the SQM Hamiltonian,
\beq
H_{\rm SQM}={p^2 \over 2 \, M_r} + \left( W^{'}(s) \right )^2 +
{W^{''}(s)
\over
\sqrt{2 \, M_r}}\, \sigma_3 ~,
\label{wqm2}
\eeq
we readily derive  the superpotential (up to a sign)
\beq
W'(s)= \sqrt{3} \, {~m^{3/2} \over \lambda} \, {\mu_2 \, \kappa\,
w_{\nu_1}(m \, s) - \nu_2 \over \kappa^{1/2} \,
(\nu_{1+}\nu_{1-})^{1/2}}~.
\label{spotqm}
\eeq
This coincides with the result (\ref{suppote}) derived in Sec.~\ref{sec:4.1},
by evaluating the field-theoretic supercharges.

\newpage

\end{document}